\documentclass[preprint]{aastex63}

\def\trx{T_\mathrm{RX}}
\def\tsys{T_\mathrm{sys}}
\def\tsky{T_\mathrm{sky}}
\def\tamb{T_\mathrm{amb}}

\def\tastar{T_\mathrm{A}^*}
\def\tcmb{T_\mathrm{CMB}}
\def\tsrc{T_\mathrm{src}}
\def\etal{\eta_\mathrm{l}}
\def\rmon{\mathrm{ON}}
\def\rmoff{\mathrm{OFF}}

\received{2020-11-12}
\revised{2020-12-30}
\accepted{2021-01-27}
\submitjournal{PASP}

\shorttitle{Correcting Atmospheric Effects on Single-Dish Spectroscopy}
\shortauthors{Sawada et al.}

\begin{document}

\title{Offline Correction of Atmospheric Effects on Single-Dish Radio Spectroscopy}

\correspondingauthor{Tsuyoshi Sawada}

\author[0000-0002-0588-5595]{Tsuyoshi Sawada}
\affiliation{Joint ALMA Observatory,
Alonso de C\'ordova 3107, Vitacura, Santiago
763-0355, Chile}
\affiliation{NAOJ Chile, National Astronomical Observatory of Japan,
Alonso de C\'ordova 3788 Office 61B, Vitacura, Santiago
763-0492, Chile}

\author{Chin-Shin Chang}
\affiliation{Joint ALMA Observatory,
Alonso de C\'ordova 3107, Vitacura, Santiago
763-0355, Chile}

\author{Harold Francke}
\affiliation{Joint ALMA Observatory,
Alonso de C\'ordova 3107, Vitacura, Santiago
763-0355, Chile}

\author{Laura Gomez}
\affiliation{Joint ALMA Observatory,
Alonso de C\'ordova 3107, Vitacura, Santiago
763-0355, Chile}

\author[0000-0003-1183-9293]{Jeffrey G.\ Mangum}
\affiliation{National Radio Astronomy Observatory,
520 Edgemont Road, Charlottesville, VA
22903, USA}

\author[0000-0002-7616-7427]{Yusuke Miyamoto}
\affiliation{National Astronomical Observatory of Japan,
2-21-1 Osawa, Mitaka, Tokyo
181-8588, Japan}

\author[0000-0003-3780-8890]{Takeshi Nakazato}
\affiliation{National Astronomical Observatory of Japan,
2-21-1 Osawa, Mitaka, Tokyo
181-8588, Japan}

\author{Suminori Nishie}
\affiliation{National Astronomical Observatory of Japan,
2-21-1 Osawa, Mitaka, Tokyo
181-8588, Japan}

\author[0000-0003-3644-807X]{Neil M.\ Phillips}
\affiliation{European Southern Observatory,
Karl-Schwarzschild-Strasse 2, Garching bei M\"unchen
85748, Germany}

\author[0000-0001-9368-3143]{Yoshito Shimajiri}
\affiliation{National Astronomical Observatory of Japan,
2-21-1 Osawa, Mitaka, Tokyo
181-8588, Japan}

\author{Kanako Sugimoto}
\affiliation{National Astronomical Observatory of Japan,
2-21-1 Osawa, Mitaka, Tokyo
181-8588, Japan}

\begin{abstract}
We present a method to mitigate the atmospheric effects (residual
atmospheric lines) in single-dish radio spectroscopy caused by
the elevation difference between the target and reference positions.
The method is developed as a script using the Atmospheric
Transmission at Microwaves (ATM) library built into the Common Astronomy
Software Applications (CASA) package.
We apply the method to the data taken with the Total Power Array of the
Atacama Large Millimeter/submillimeter Array.
The intensities of the residual atmospheric (mostly O$_3$) lines are
suppressed by, typically, an order of magnitude for the tested cases.
The parameters for the ATM model can be optimized to minimize the
residual line and, for a specific O$_3$ line at 231.28 GHz,
a seasonal dependence of a best-fitting model parameter is demonstrated.
The method will be provided as a task within the CASA package
in the near future.
The atmospheric removal method we developed can be used by any
radio/millimeter/submillimeter observatory to improve the quality of
its spectroscopic measurements.
\end{abstract}

\keywords{
\href{http://astrothesaurus.org/uat/1861}{Astronomy data reduction (1861)} ---
\href{http://astrothesaurus.org/uat/1061}{Millimeter astronomy (1061)} ---
\href{http://astrothesaurus.org/uat/1359}{Radio spectroscopy (1359)} ---
\href{http://astrothesaurus.org/uat/1647}{Submillimeter astronomy (1647)}
}

\section{Introduction}\label{sec:intro}

The chopper-wheel method
\citep[e.g.,][]{1976ApJS...30..247U,1981ApJ...250..341K} is commonly used
for calibrating the amplitude of the data from single-dish radio telescopes,
\begin{equation}
\tastar = \frac{P(\rmon)-P(\rmoff)}{P(\rmoff)} \tsys ~~, \label{eq:chopperwheel}
\end{equation}
where $\tastar$ is the antenna temperature corrected for the atmospheric
attenuation and antenna losses, $\tsys$ is the system noise temperature,
$P(\rmon)$ and $P(\rmoff)$ are the detector readouts toward the target (ON)
and astronomical-emission-free reference (OFF) positions, respectively
(a detailed formulation is given in Section \ref{sec:method}).

A fundamental assumption in Equation (\ref{eq:chopperwheel}) is that the
atmospheric emission/absorption is the same between the ON and OFF positions.
When this assumption breaks-down, the subtraction of the sky emission in
the ON and OFF measurements becomes imperfect and produces artifacts in
the resultant calibrated data.
A common cause of the non-optimal subtraction of the sky emission is the
elevation difference between the ON and OFF positions.
For isolated and compact targets like external galaxies, the difference
can be minimized by acquiring the OFF position at the same elevation as
the ON position.
However such a mitigation is not possible for extended objects like
the Galactic molecular clouds, which require carefully-chosen OFF positions,
which are known to be free of source emission (i.e., CO in Galactic
molecular clouds), at fixed celestial coordinates.
In the case of the Atacama Large Millimeter/submillimeter Array (ALMA)
Total Power Array \citep[TP Array;][]{2009PASJ...61....1I} observations
of such objects, the separation between the ON and OFF positions can be
as large as a few degrees.

\begin{figure}
\plotone{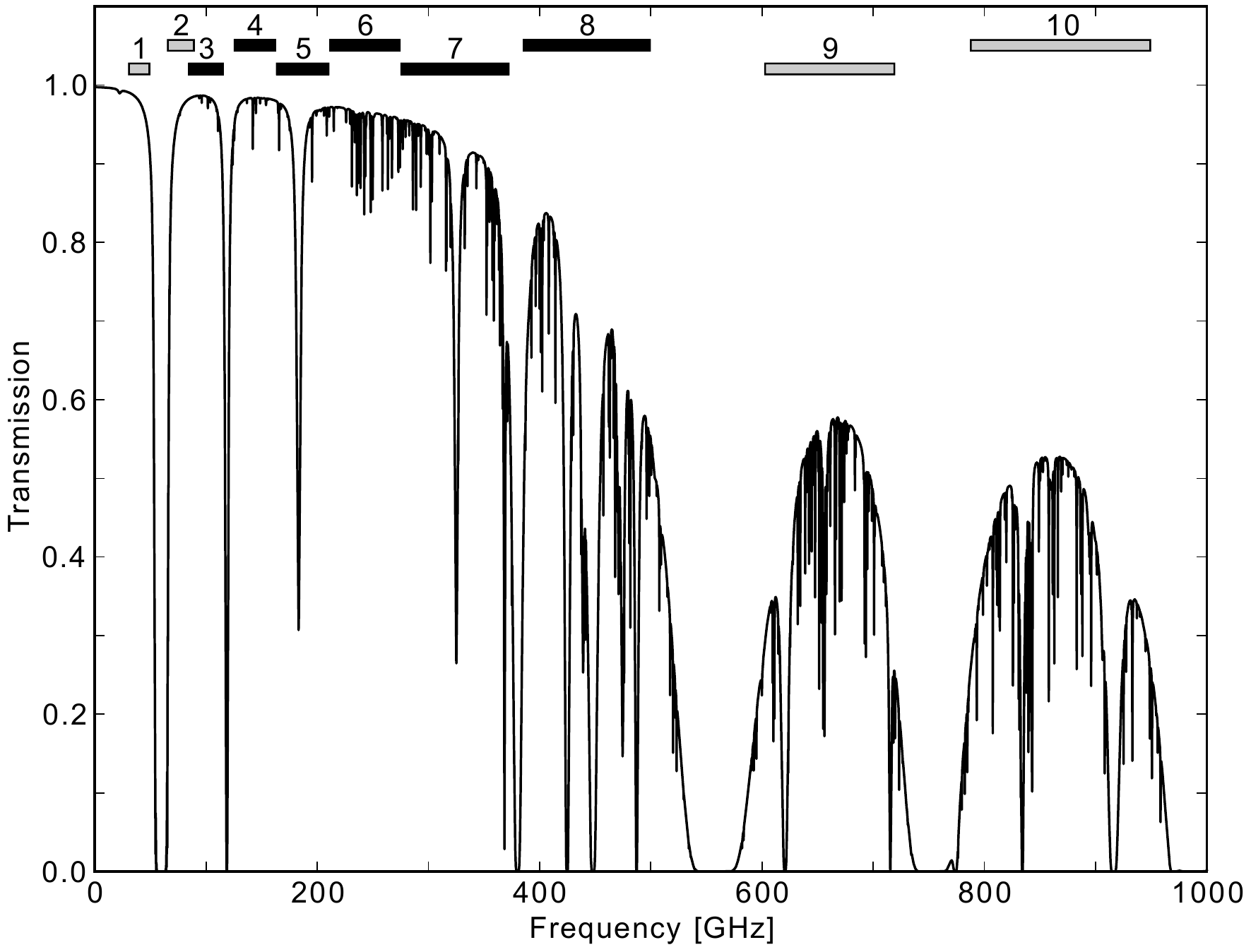}
\caption{The transmission of an atmospheric model at the ALMA site toward
zenith (0.5 mm precipitable water vapor is assumed) computed by CASA/ATM
(see Section \ref{sec:implementation}).
The horizontal bars indicate the frequency coverages of the ALMA
receiver bands.
The bands that are currently offered for TP Array observations are
drawn in black, while others are in gray.
\label{fig:transmission}}
\end{figure}

As long as the atmospheric emission and absorption can be considered to be
quasi-continuum within the instantaneous bandwidth of the
spectrometer\footnote{In this article, we are concerned with spectral line
observations only.}, the artifact is also quasi-continuum.
It can be easily corrected, to the first order\footnote{To be more precise,
the amplitude scaling also needs to be corrected, see
Section \ref{sec:method}, Equation (\ref{eq:scale}).},
by subtracting spectral baselines.
However, that is no longer the case if relatively narrow lines from
atmospheric species fall into the measured bandwidth.
This issue is becoming more critical for mm and sub-mm observations
due to the broad instantaneous bandwidths which modern instruments
offer,
given the existence of multiple telluric lines from species such as
H$_2$O, O$_2$, and O$_3$
(Figure \ref{fig:transmission}).

In this article, we present a method for offline
(i.e., post-processing) correction of the atmospheric artifacts.
The method itself is formulated in Section \ref{sec:method}.
We implemented the method within the Common Astronomy Software Applications
package \citep[CASA;][]{2007ASPC..376..127M} as described in
Section \ref{sec:implementation}.
The verification of the method using the data from the ALMA
TP Array is presented in Section \ref{sec:results}.

\section{Method}\label{sec:method}

The detector (spectrometer) readouts toward the ON and OFF positions are,
assuming the ideal single-sideband response\footnote{The typical sideband
rejection of the ALMA sideband-separating receivers (Bands 3--8)
is 20 dB (W.\ Dent, private communication).},
written as
\begin{eqnarray}
P(\rmon) &=& G k \left\{ \trx + \etal \tsky(\rmon) + \etal e^{-\tau(\rmon)} \tcmb + (1-\etal) \tamb + \etal e^{-\tau(\rmon)}\tsrc \right\} \\
P(\rmoff) &=& G k \left\{ \trx + \etal \tsky(\rmoff) + \etal e^{-\tau(\rmoff)} \tcmb + (1-\etal) \tamb \right\}
\end{eqnarray}
where
$G$ is the system gain;
$k$ is the Boltzmann constant;
$\trx$ is the receiver noise temperature;
$\etal$ is the feed efficiency \citep[$1-\eta_{\rm l}$ corresponds to
  the rearward spillover, blockage, and ohmic losses;][]{1976ApJS...30..247U};
$\tsky$ is the radiation temperature of the sky toward the line of sight;
$\tau$ is the optical depth of the atmosphere toward the line of sight;
$\tcmb$ is the radiation temperature of the cosmic microwave background (CMB);
$\tamb$ is the ambient temperature;
$\tsrc$ is the radiation temperature of the target object.
All the variables are functions of frequency although not explicitly written,
and all of the temperatures are Rayleigh--Jeans equivalent temperatures.
Note that $\tsky$ includes the effect of the atmospheric opacity;
it cannot simply be expressed as $\mathrm{constant}\times (1-e^{-\tau})$
since the temperature of the atmosphere varies with altitude.

The calibrated antenna temperature is 
\begin{eqnarray}
\tastar &=& \frac{P(\rmon)-P(\rmoff)}{P(\rmoff)} \tsys \nonumber \\
&=& G k \etal \left\{ \tsky(\rmon)-\tsky(\rmoff) + \left( e^{-\tau(\rmon)}-e^{-\tau(\rmoff)} \right) \tcmb + e^{-\tau(\rmon)}\tsrc \right\} \frac{\tsys}{P(\rmoff)}
\end{eqnarray}
and the system noise temperature is
\begin{equation}
\tsys = \frac{P e^\tau}{G k \etal} ~~,
\end{equation}
hence\footnote{In ALMA TP Array observations,
$\tsys$ is measured at the OFF position.
Although $\tsys$ is not measured concurrently
with OFF or ON measurements,
the time-interpolated $\tsys$ is used for
calibration into individual $\tastar$ spectra.
Therefore $\tsys \approx P(\rmoff)\exp(\tau(\rmoff))/(Gk\etal)$.}
\begin{equation}
\tastar \approx \left\{ \tsky(\rmon)-\tsky(\rmoff) + \left( e^{-\tau(\rmon)}-e^{-\tau(\rmoff)} \right) \tcmb + e^{-\tau(\rmon)}\tsrc \right\} e^{\tau(\rmoff)} \label{eq:tareal}
\end{equation}
while it should ideally be
\begin{equation}
\tastar = \tsrc ~~. \label{eq:taideal}
\end{equation}

Therefore, in order to mitigate the artifacts caused by the difference
between the atmosphere toward the ON and OFF positions,
the difference between Equations (\ref{eq:tareal}) and (\ref{eq:taideal})
\begin{eqnarray}
{\mit\Delta}\tastar &\approx& \left\{ \tsky(\rmon)-\tsky(\rmoff) + \left( e^{-\tau(\rmon)}-e^{-\tau(\rmoff)} \right) \tcmb \right\} e^{\tau(\rmoff)} + \left( e^{\tau(\rmoff)-\tau(\rmon)}-1 \right) \tsrc  \label{eq:dtaall} \nonumber \\
&\equiv& {\mit\Delta}\tastar(1) + {\mit\Delta}\tastar(2)
\end{eqnarray}
needs to be computed using atmospheric models (e.g., the millimeter-wave
propagation model [MPM] by \citealt{1989IJIMW..10..631L}; the atmospheric
transmission at microwaves [ATM] by \citealt{2001ITAP...49.1683P}) and
subtracted from the calibrated spectra.
The first term corresponds to the residual atmospheric lines (plus
quasi-continuum, which is eliminated by spectral baseline subtraction),
while the second term indicates that $\tsrc$ is incorrectly scaled
by a factor of $\exp({\tau(\rmoff)-\tau(\rmon)})$.

The second term ${\mit\Delta}\tastar(2)$ is not directly calculable,
since $\tsrc$ is unknown.
Instead of subtracting ${\mit\Delta}\tastar(2)$ from $\tastar$,
the correction can be made by
\begin{eqnarray}
\tastar(\mathrm{corrected}) &=& \left\{ \tastar-{\mit\Delta}\tastar(1) \right\} e^{\tau(\rmon)-\tau(\rmoff)} \nonumber \\
&=& \tsrc  ~~. \label{eq:scale}
\end{eqnarray}

If we assume the atmospheric transmission of 95\% at the quasi-continuum,
90\% at the line (typical values for O$_3$ lines in ALMA Band 6
[211--275 GHz] observations; Figure \ref{fig:transmission}), and
the elevations of the ON and OFF positions of $50\arcdeg$ and $49\arcdeg$,
respectively, ${\mit\Delta}\tastar(1)$ amounts to $\simeq 0.2\;\mathrm{K}$
after re-subtraction of a spectral baseline.
On the other hand, ${\mit\Delta}\tastar(2)$ is about 0.1\% and 0.2\% of
$\tsrc$, respectively, at the quasi-continuum and line frequencies.
That is, ${\mit\Delta}\tastar(2)$ is smaller than ${\mit\Delta}\tastar(1)$,
unless $\tsrc$ is extremely high ($\gtrsim 100\;\mathrm{K}$) in that condition.
It is also much smaller than the absolute flux accuracy ALMA currently offers,
5\% in Bands 3 to 5 and 10\% in Bands 6 to 8 \citep{ALMATHCy8}.
Although ${\mit\Delta}\tastar(2)$ may become significant under
some circumstances (in particular at low elevations), we have not been
able to collect sufficient cases to validate it.
Therefore, we focus on the correction expressed by ${\mit\Delta}\tastar(1)$
in this article and defer the validation of ${\mit\Delta}\tastar(2)$
to a future study.

\section{Implementation}\label{sec:implementation}

In order to verify the method, we wrote a script in CASA (version 5.6.1)
which calculates the first term of the right-hand side of Equation (\ref{eq:dtaall})
\begin{eqnarray}
{\mit\Delta}\tastar(1) &\equiv& \left\{ \tsky(\rmon)-\tsky(\rmoff) + \left( e^{-\tau(\rmon)}-e^{-\tau(\rmoff)} \right) \tcmb \right\} e^{\tau(\rmoff)}  \label{eq:dtaline}
\end{eqnarray}
and subtracts it from the calibrated spectra ($\tastar$).

We use the ATM model \citep{2001ITAP...49.1683P} to compute the radiation
temperature and opacity of the atmosphere.
The ATM library is integrated into CASA and accessible via its
{\tt atmosphere} Toolkit functions.
First, the altitude of the telescope and the atmospheric conditions
(i.e., the temperature, pressure, and relative humidity at the ground level,
and the precipitable water vapor [PWV]) are read from the metadata of the
observations and passed to the {\tt initAtmProfile} and {\tt setUserWH2O}
functions.
Second, the spectral setup (i.e., the center frequency, channel spacing, and
the total bandwidth of each spectral window) is also read from the metadata
and used to invoke the {\tt initSpectralWindow} function.
Finally, the radiation temperature and opacity of the sky are obtained by
the {\tt getTrjSkySpec}, {\tt getDryOpacitySpec}, and {\tt getWetOpacitySpec}
functions for every ON and OFF position, whose airmass (elevation) is read
from the metadata and set to the {\tt setAirMass} function.

The radiation temperature of the sky calculated by the {\tt getTrjSkySpec}
function, $\tsky^\mathtt{ATM}$, already includes the CMB term (unless the
CMB temperature is manually set to 0).
That is,
\begin{eqnarray}
\tsky^\mathtt{ATM} &=& \tsky + e^{-\tau}\tcmb
\end{eqnarray}
and, therefore, Equation (\ref{eq:dtaline}) can be rewritten into a simpler form
\begin{eqnarray}
{\mit\Delta}\tastar(1) &=& \left\{ \tsky^\mathtt{ATM}(\rmon)-\tsky^\mathtt{ATM}(\rmoff) \right\} e^{\tau(\rmoff)} ~~. \label{eq:dtaline2}
\end{eqnarray}
Additionally, the spectral response of the spectrometer is convolved with
the ${\mit\Delta}\tastar(1)$ spectra if necessary.

Although some of the parameters for the {\tt initAtmProfile} function are
``fixed'' as described above, the function also accepts some other ``free''
parameters, such as the atmosphere type ({\tt atmType}, corresponding to
the latitude and season), the altitude of the top of the modeled atmosphere
({\tt maxAltitude}), the lapse rate ({\tt dTem\_dh}), and the scale height
of water vapor ({\tt h0}).
The available parameters and their default values are summarized in
Table \ref{tab:parameters}.
In the next section we explore what the best parameters are for
ALMA TP Array data.

\begin{deluxetable*}{llr}
\tablecolumns{3}
\tablewidth{0pc}
\tablecaption{Parameters for the {\tt initAtmProfile} function\label{tab:parameters}}
\tablehead{
  \colhead{Name} & \colhead{Explanation} & \colhead{Default value}
}
\startdata
{\tt altitude} & Site altitude & 5000 m\tablenotemark{a}\\
{\tt temperature} & Ambient temperature & 270 K\tablenotemark{a}\\
{\tt pressure} & Ambient pressure & 560 hPa\tablenotemark{a}\\
{\tt maxAltitude} & Altitude of the top of the modelled atmosphere & 48 km\tablenotemark{b}\\
{\tt humidity} & Humidity & 20 \%\tablenotemark{a}\\
{\tt dTem\_dh} & Derivative of temperature with respect to height & $-5.6$ $\mathrm{K\,km^{-1}}$\\
{\tt dP} & Initial pressure step & 10 hPa\\
{\tt dPm} & Pressure multiplicative factor for steps & 1.2\\
{\tt h0} & Scale height for water & 2 km\\
{\tt atmType} & Atmospheric type & 1\tablenotemark{c}\\
{\tt layerBoundaries} & Altitude of user-defined temperature profile & ---\\
{\tt layerTemperature} & User-defined temperature profile & ---\\
\enddata
\tablenotetext{a}{We use the values read from the metadata (see 
  Section \ref{sec:implementation}).}
\tablenotetext{b}{We use 120 km instead of the default value in this
  article, unless otherwise noted (see Section \ref{sec:parameters}).}
\tablenotetext{c}{Enumerated to 1 ({\it tropical\/}),
  2 ({\it mid latitude summer\/}), 3 ({\it mid latitude winter\/}),
  4 ({\it subarctic summer\/}), and 5 ({\it subarctic winter\/}).}
\end{deluxetable*}%

\section{Results and Discussion}\label{sec:results}

We apply the method described above to a selection of archival data from
the ALMA TP Array and present the results.
The data are calibrated into $\tastar$ through the standard way
(Equation \ref{eq:chopperwheel}; using CASA's {\tt sdcal} task) first, and
then ${\mit\Delta}\tastar(1)$ (Equation \ref{eq:dtaline2}) is subtracted.

\subsection{Proof of Concept}

\begin{figure}
\plotone{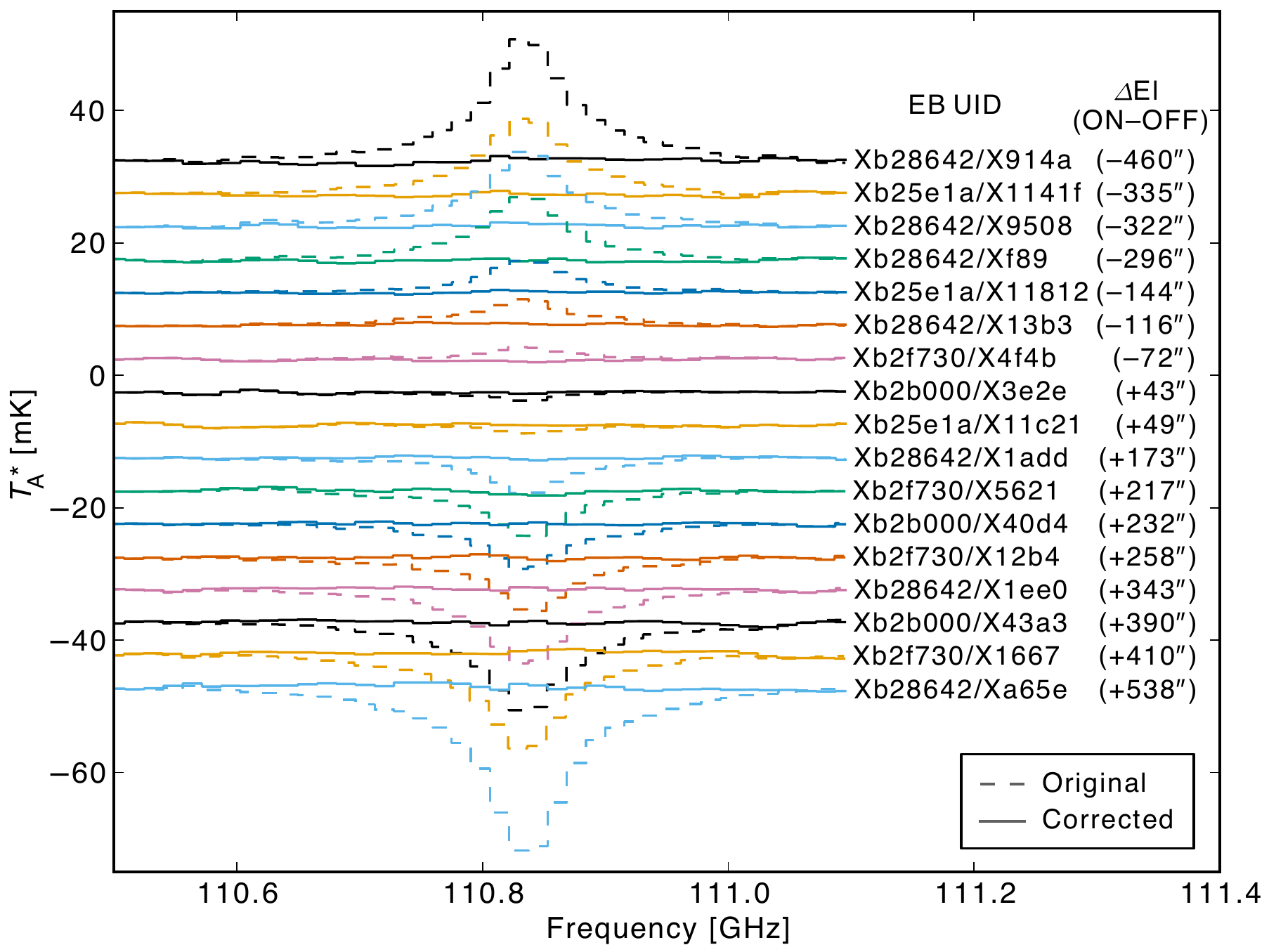}
\caption{The averaged $\tastar$ spectra (+offsets) of the 110.84 GHz
O$_3$ line from 17 EBs for the SB Thakeray\_a\_03\_TP.
The observations were made in 2016 May.
The dashed and solid lines are the spectra before and after correction,
respectively.
The spectra are sorted by the hour angles of the observations.
The unique IDs (UIDs) of the EBs are annotated on the right (the common
prefix, ``uid://A002/'', is omitted), along with the median elevation
differences between the ON (map center) and OFF positions in the parentheses.
The {\it mid latitude summer\/} model (${\tt atmType}=2$) is used for
the correction.
\label{fig:thakeray}}
\end{figure}

\begin{figure}
\plotone{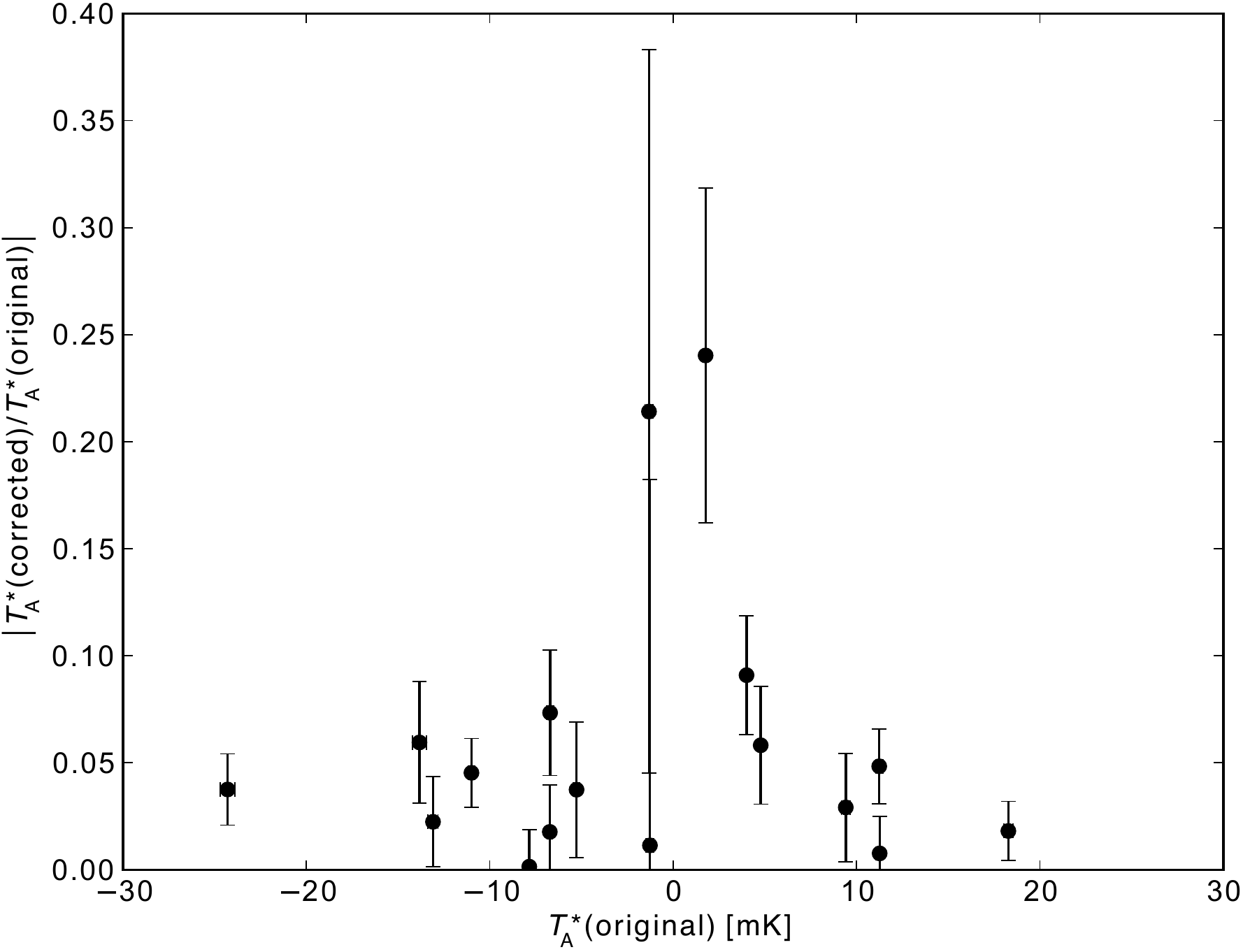}
\caption{The ratios between the corrected and original $\tastar$,
at the center frequency of the line, for the spectra shown in
Figure \ref{fig:thakeray}.
The error bars indicate the $1\sigma$ noise.
\label{fig:thakeray_suppressionratio}}
\end{figure}

Figure \ref{fig:thakeray} shows an example in ALMA Band 3; the averaged
$\tastar$ spectra before and after correction from seventeen
execution blocks (EBs) for the scheduling block (SB) Thakeray\_a\_03\_TP
of the project 2015.1.00908.S.
The OFF position was approximately at $-11\arcmin$ toward the
R.A.\ direction from the field center.
Therefore, during these measurements, the ON position was at lower and
higher elevation than the OFF position before and after the transit,
respectively.
The spectra before correction demonstrate that the amplitude of the
residual O$_3$ line at 110.84 GHz clearly correlates with the
elevation difference between ON and OFF, as expected.
The line is mostly eliminated in the spectra after the correction
(i.e., subtracting ${\mit\Delta}\tastar(1)$).
The ratios between the corrected and original $\tastar$ of the line are
displayed in Figure \ref{fig:thakeray_suppressionratio}.
At the frequency resolution of these measurements (31 MHz),
the line intensities are suppressed by more than an order of magnitude,
excepting a few that are noise-dominated (small $\tastar$ without
correction, i.e., observed around the transit).

\begin{figure}
\plotone{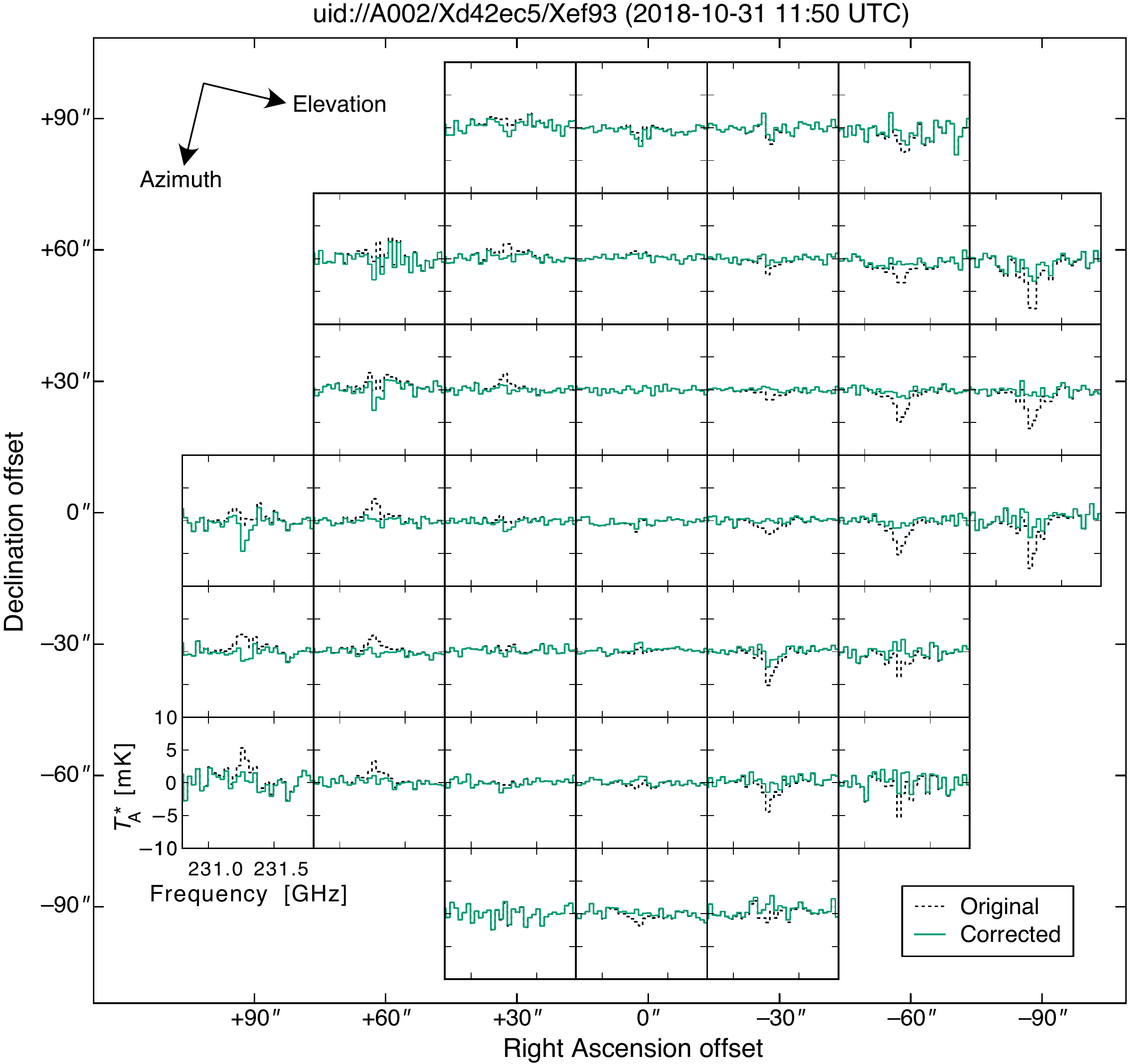}
\caption{The 231.28 GHz O$_3$ line profile map made from the EB UID
uid://A002/Xd42ec5/Xef93 (SB NGC4038\_a\_06\_TP, project 2018.1.00272.S).
Each panel represents a $30\arcsec\times 30\arcsec$ area.
The dashed and solid lines are the spectra before and after correction
(with the {\it mid latitude summer\/} model), respectively.
The equatorial coordinates are shown as offsets with respect to
$(\mathrm{R.A.}, \mathrm{Decl.})_\mathrm{J2000} =
(12^\mathrm{h} 01^\mathrm{m} 53\fs 1, -18\arcdeg 52\arcmin 32\arcsec)$.
The directions of the horizontal axes (azimuth--elevation) at the time
of the execution are drawn at top left.
The data have been smoothed and resampled to 31 MHz resolution/spacing
to improve the signal-to-noise ratio.
\label{fig:profmap}}
\end{figure}

As mentioned in Section \ref{sec:intro}, the elevation difference between
the ON and OFF positions can be minimized by using OFF positions that are
at the same elevation as their respective ON positions for isolated and
compact objects.
However, in practice, it is not always possible to keep the elevations
of the ON and OFF positions exactly the same, due to, e.g.,
the diurnal motion and the strategy of mapping observations.
Therefore the correction is also worth applying for such cases.
Figure \ref{fig:profmap} shows the profile map of the 231.28 GHz O$_3$
line in ALMA Band 6 before and after correction.
During the measurement which employed the On-The-Fly mapping technique
\citep{2007A&A...474..679M,2008PASJ...60..445S},
the OFF position was taken at constant elevation $+10\arcmin$ toward
the azimuthal direction from the field center, rather than the 
individual ON positions.
The spatial gradient of the residual O$_3$ line along the elevation axis
(at the position angle of $-103\arcdeg$ with respect to the equatorial
coordinates) seen in the uncorrected (original) data is mostly
eliminated from the corrected data.

\subsection{Parameter Selection}\label{sec:parameters}

\begin{figure}
\gridline{\leftfig{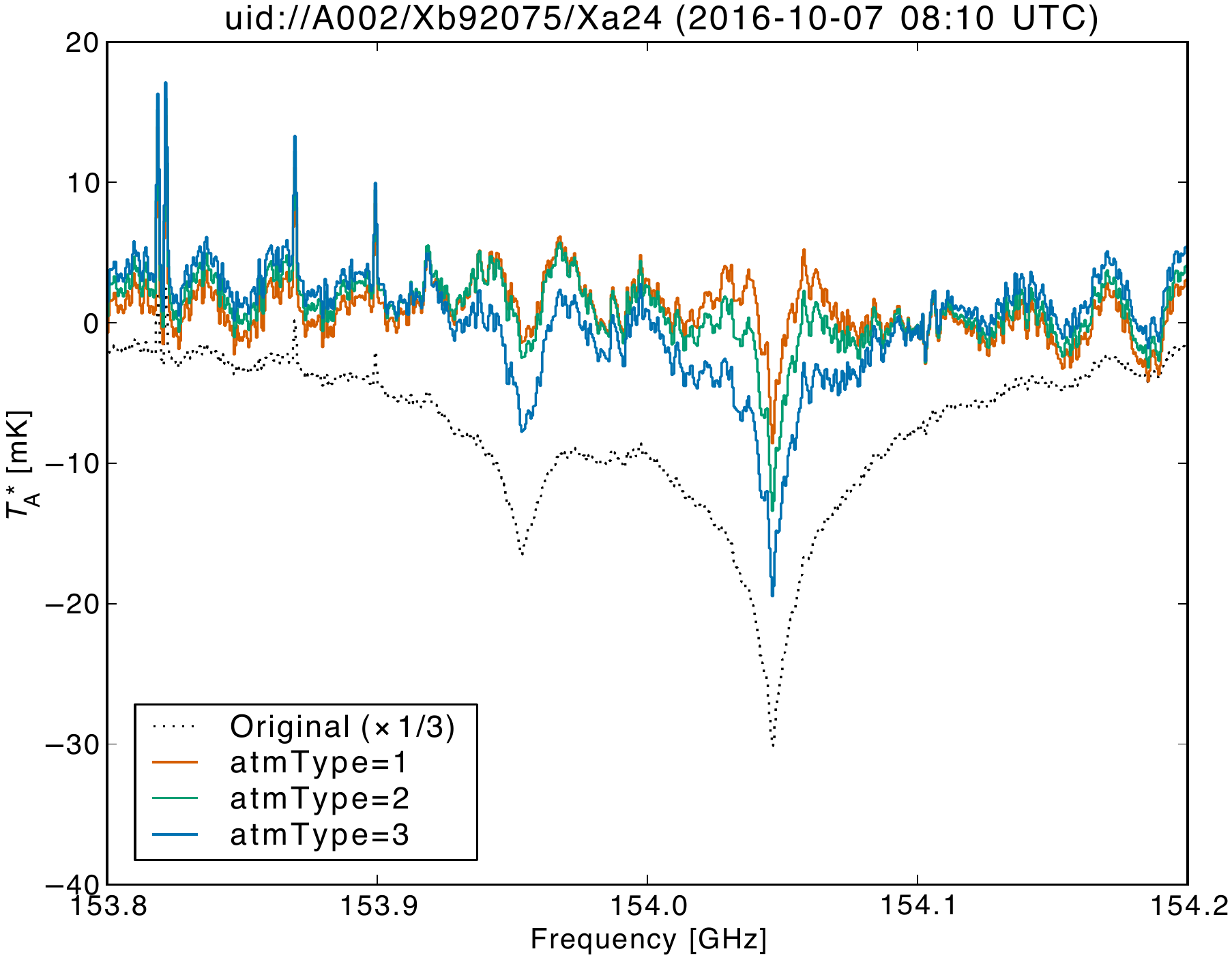}{0.5\textwidth}{(a)}
          \rightfig{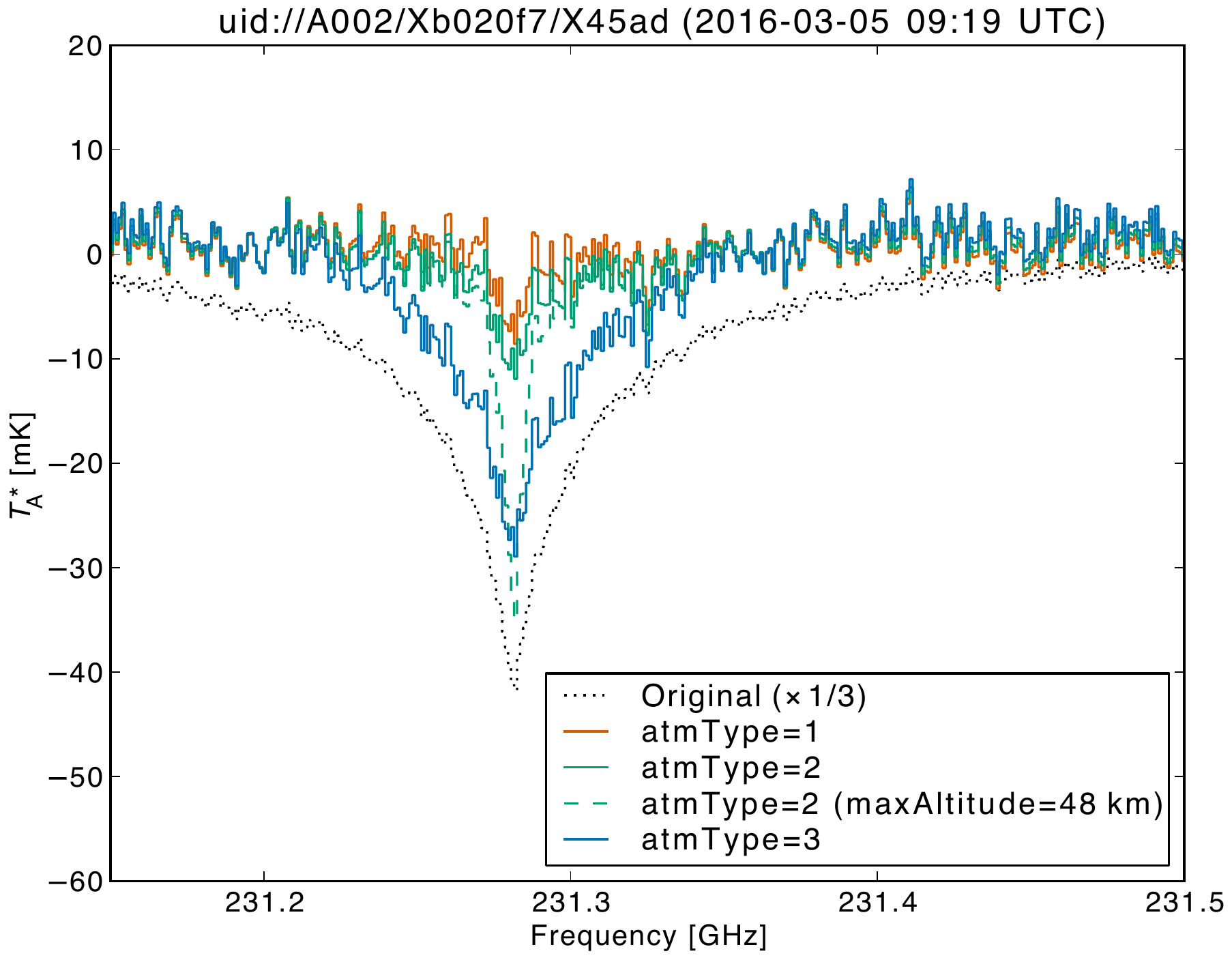}{0.5\textwidth}{(b)}}
\gridline{\leftfig{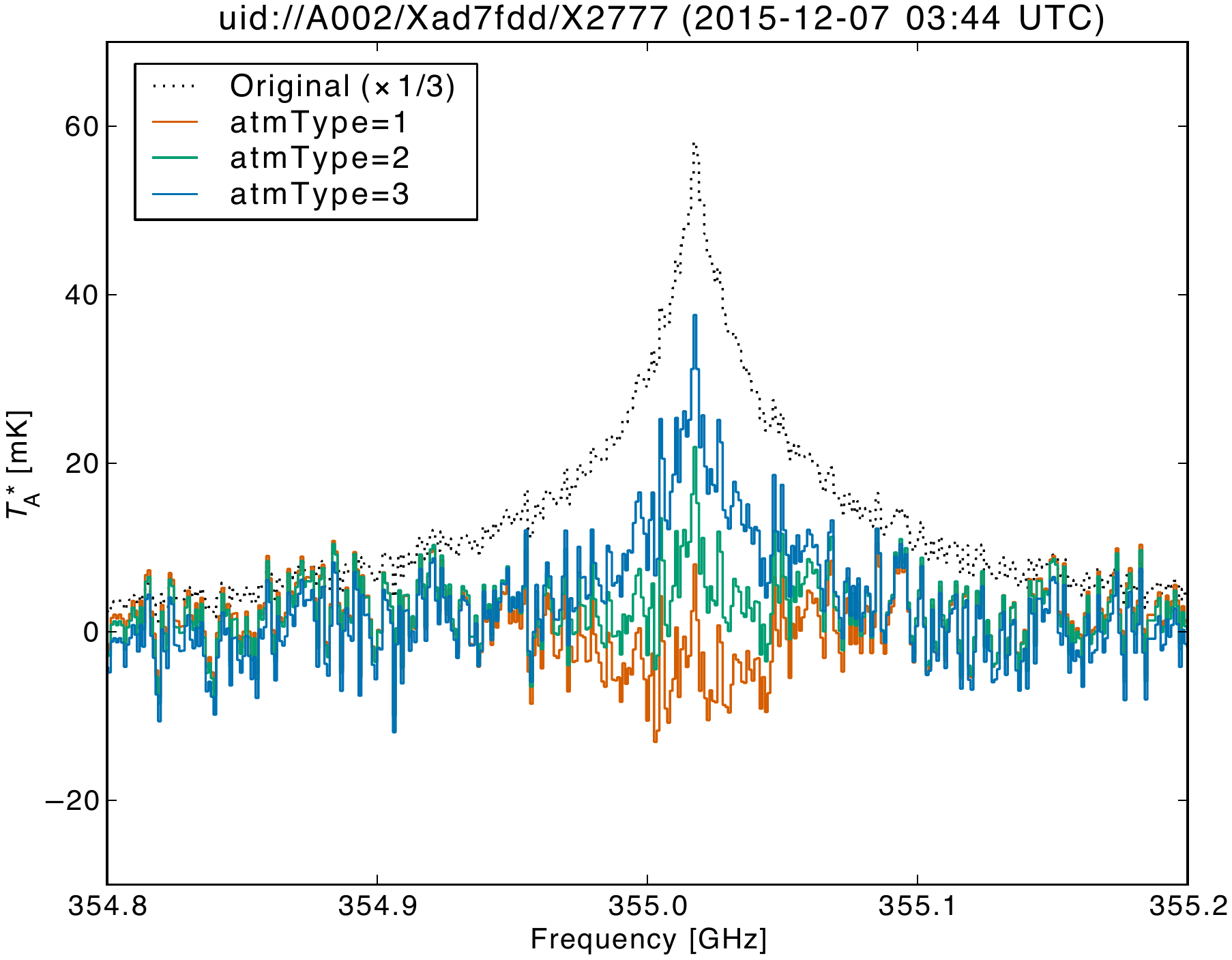}{0.5\textwidth}{(c)}
          \rightfig{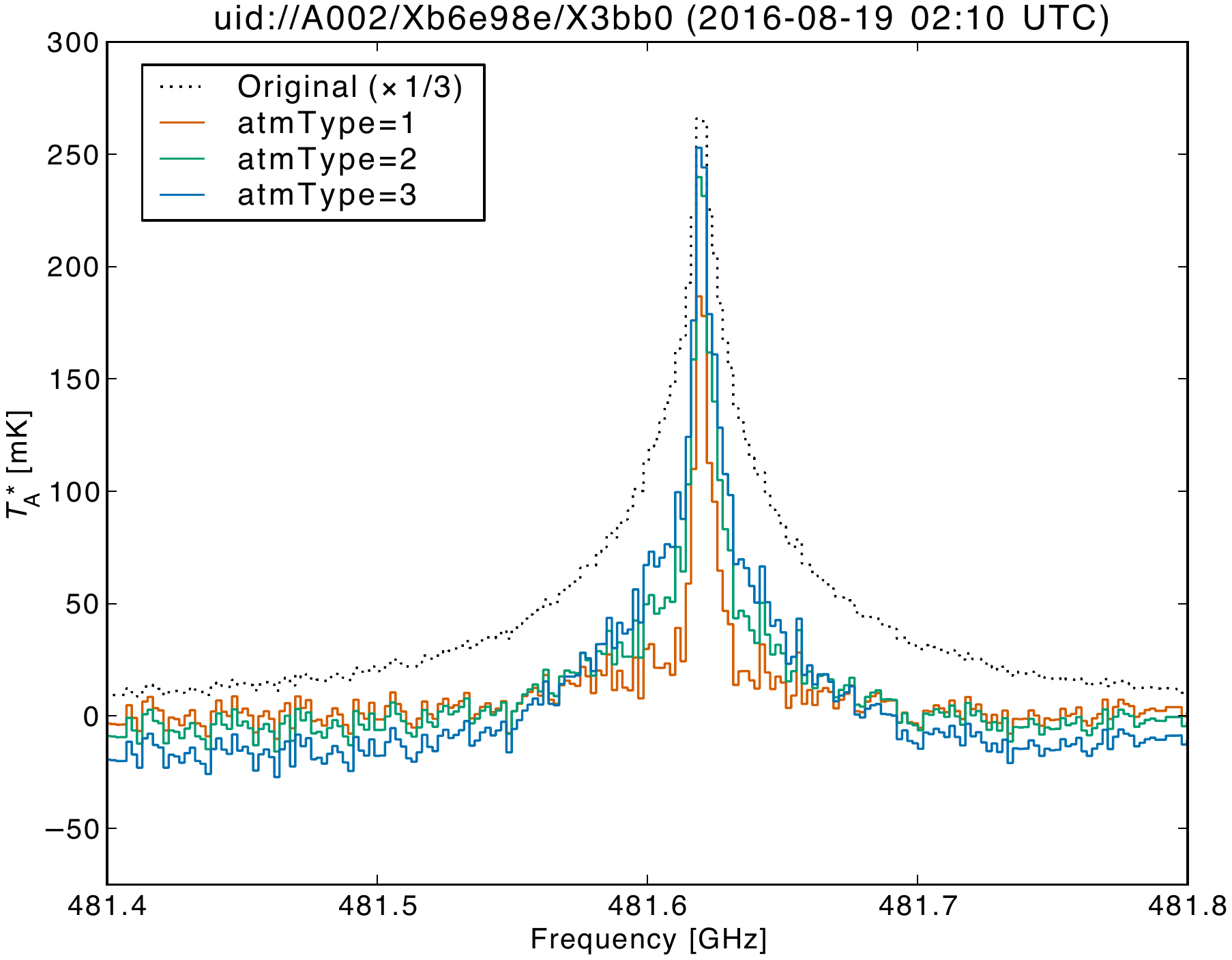}{0.5\textwidth}{(d)}}
\caption{The uncorrected and corrected spectra of the O$_3$ lines in
(a) ALMA Band 4 (EB uid://A002/Xb92075/Xa24, SB L1527\_a\_04\_TP,
  project 2016.1.01541.S),
(b) Band 6 (EB uid://A002/Xb020f7/X45ad, SB M83\_a\_06\_TP,
  project 2015.1.00121.S),
(c) Band 7 (EB uid://A002/Xad7fdd/X2777, SB NGC253\_a\_07\_TP,
  project 2015.1.00274.S), and
(d) Band 8 (EB uid://A002/Xb6e98e/X3bb0, SB NGC\_6240\_a\_08\_TP,
  project 2015.1.00717.S).
The date/time shown at the top of each panel is the centroid of
the on-source integrations.
The uncorrected spectra (dotted lines) are divided by 3 so that
they fit into the axes scale.
The corrected spectra using three {\tt atmType} values with
${\tt maxAltitude}=120\;[\mathrm{km}]$ are drawn as solid lines;
${\tt atmType}=1$ ({\it tropical\/}), 2 ({\it mid latitude summer\/}),
and 3 ({\it mid latitude winter\/}).
For the Band 6 data (top right), the corrected spectrum using
${\tt atmType}=2$, ${\tt maxAltitude}=48\;[\mathrm{km}]$ is also drawn
as dashed line.
\label{fig:otherbands}}
\end{figure}

A few more O$_3$ line profiles at higher spectral resolutions are
presented in Figure \ref{fig:otherbands}:
154.05 GHz in Band 4,
231.28 GHz in Band 6,
355.02 GHz in Band 7, and
481.62 GHz in Band 8.
Three {\tt atmType} values, out of the five available
options\footnote{The other two are for subarctic regions and therefore
should be unsuited for the data from ALMA TP Array, located at
latitude $23\arcdeg$ S.}, are attempted to correct the residual lines.
The {\tt atmType} parameter controls the vertical profiles of the
temperature and pressure of upper atmosphere and the distribution of
the molecular species.
Although none of the models can completely eliminate the line,
the {\it tropical\/} model (${\tt atmType}=1$) tends to produce
the best results in these cases, and the {\it mid latitude summer\/} model
(${\tt atmType}=2$) is slightly worse.
The best model may, however, be dependent on the observing seasons,
as discussed in Section \ref{sec:season}, as well as the species
and transitions.
The peak line intensities are suppressed by factors of
$\approx 10\mbox{--}20$, if the best models are used, for the
154.05, 231.28, and 355.02 GHz lines.
For the 481.62 GHz line, the suppression of the peak intensity is about
a factor of 4.

\begin{figure}
\gridline{\fig{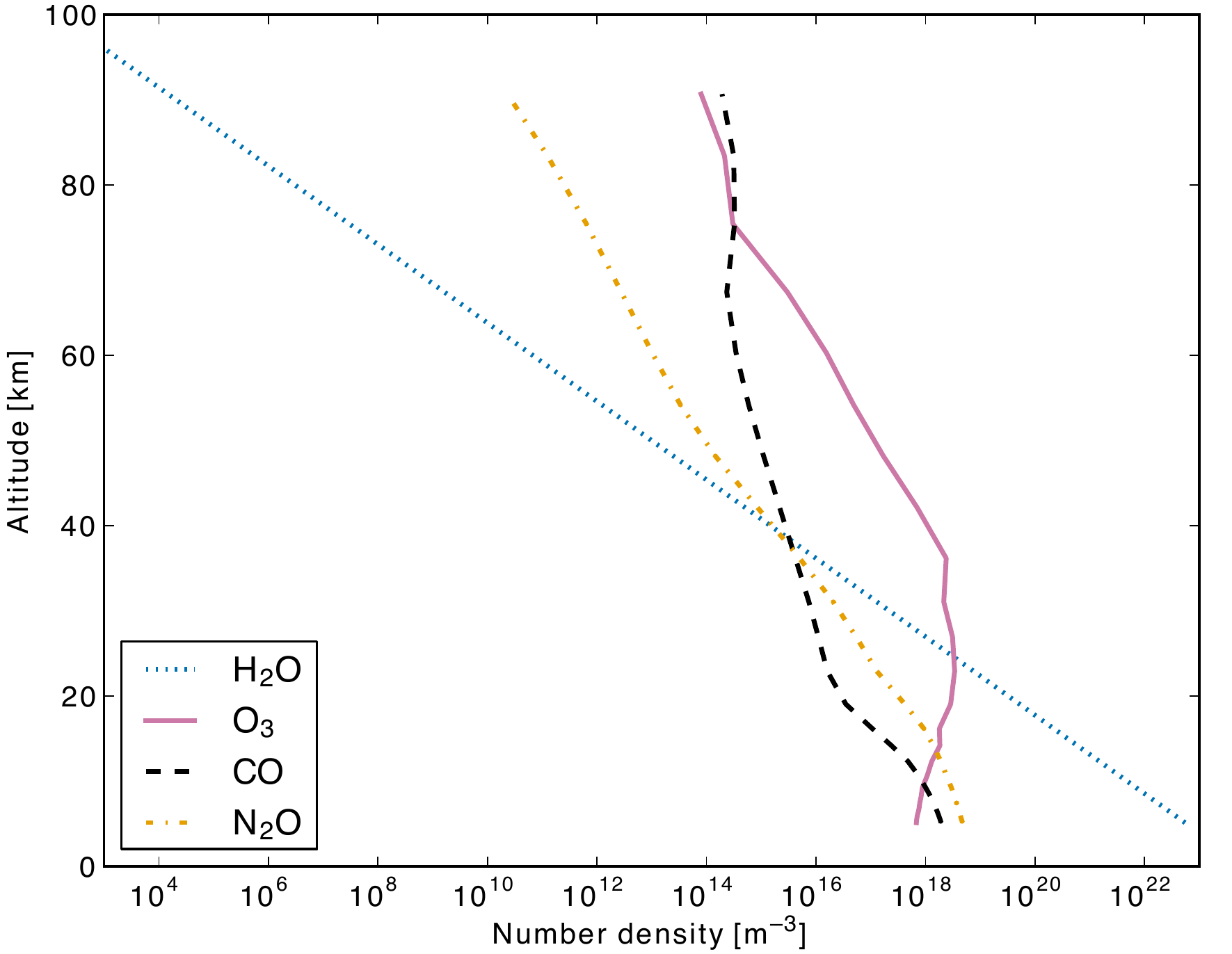}{0.7\textwidth}{(a)}}
\gridline{\fig{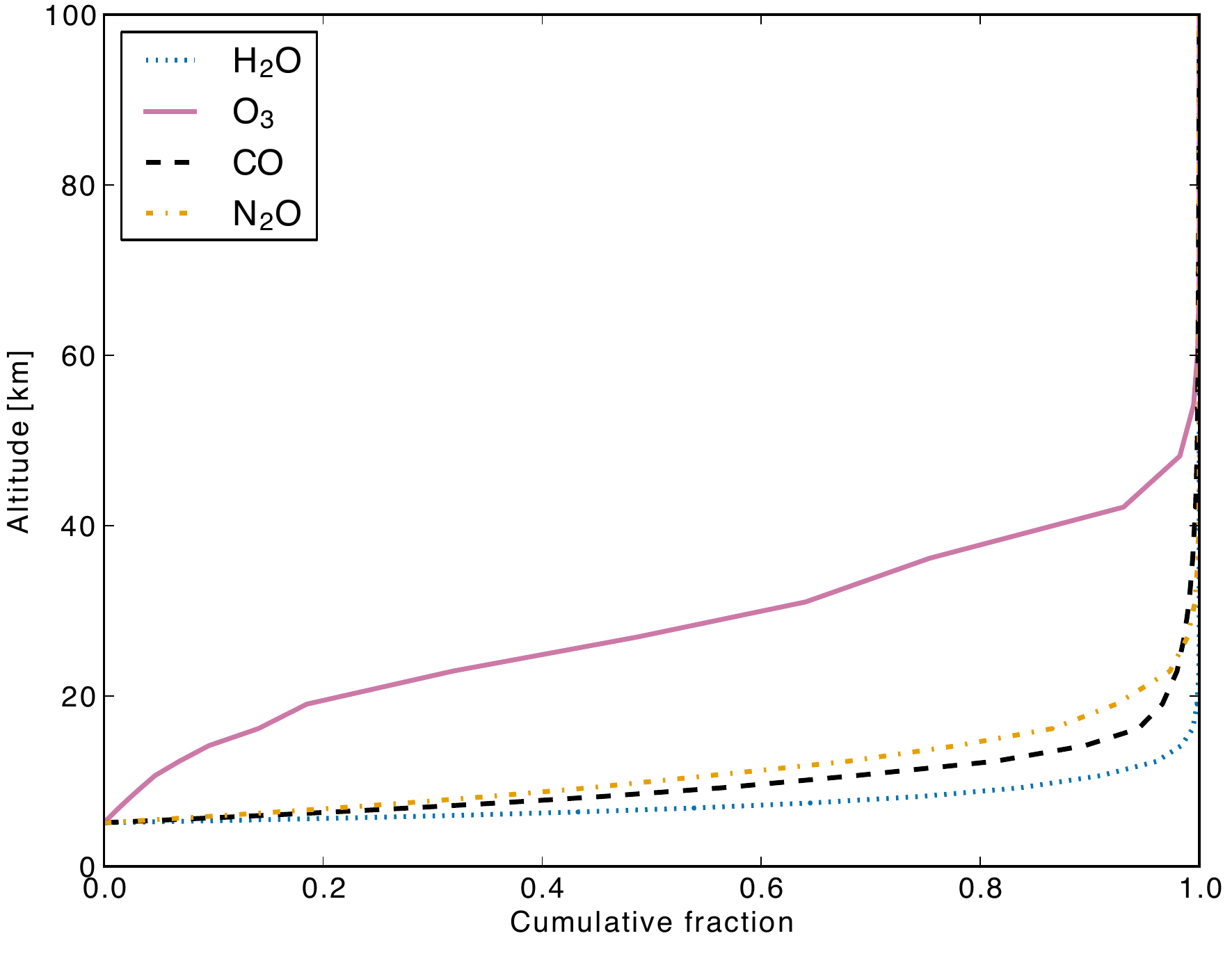}{0.7\textwidth}{(b)}}
\caption{(a) The number densities and (b) cumulative column density
fractions of the H$_2$O, O$_3$, CO, and N$_2$O molecules as a function
of altitude above sea level.
Computed by CASA/ATM using the {\it mid latitude summer\/}
(${\tt atmType}=2$) model for the atmospheric conditions of 268 K
ambient temperature, 556 hPa pressure, 50\% humidity, and 1.4 mm PWV
(close to the conditions for uid://A002/Xb020f7/X45ad,
Figure \ref{fig:otherbands}b).
\label{fig:atmprof}}
\end{figure}

The {\tt maxAltitude} of 120 km  was used to calculate the corrected
spectra in Figure \ref{fig:otherbands}.
In addition, for the 231.28 GHz line (panel b), the corrected
spectrum using $\mathtt{maxAltitude}=48\;\mathrm{km}$, which is the
default value of the {\tt initAtmProfile} function, is also drawn.
It demonstrates that the 48 km {\tt maxAltitude} is not sufficient,
leaving a sharp residual at the center frequency of the line.
Figure \ref{fig:atmprof} shows the computed number densities of the
H$_2$O, O$_3$, CO, and N$_2$O molecules and their cumulative fraction
as a function of altitude.
A few \% of the O$_3$ molecules are distributed above 48 km.
Although being a small fraction compared with the bulk at
lower altitude, it results in a remarkably deficient
correction of the line due to the reduced pressure
broadening, as shown above.
Therefore $\mathtt{maxAltitude}=120\;\mathrm{km}$ is used in this
article, unless otherwise noted.

\begin{figure}
\gridline{\leftfig{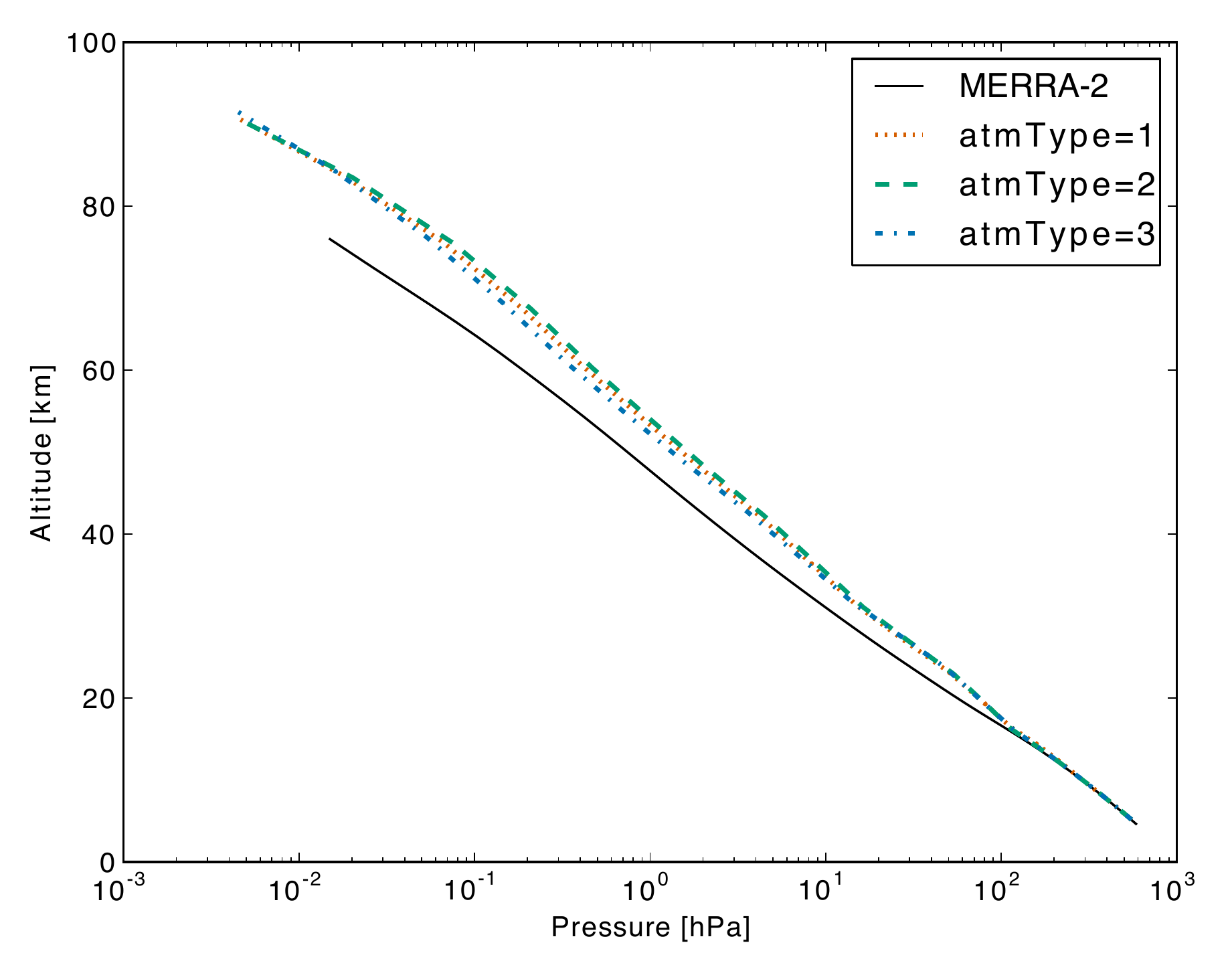}{0.5\textwidth}{(a)}
          \rightfig{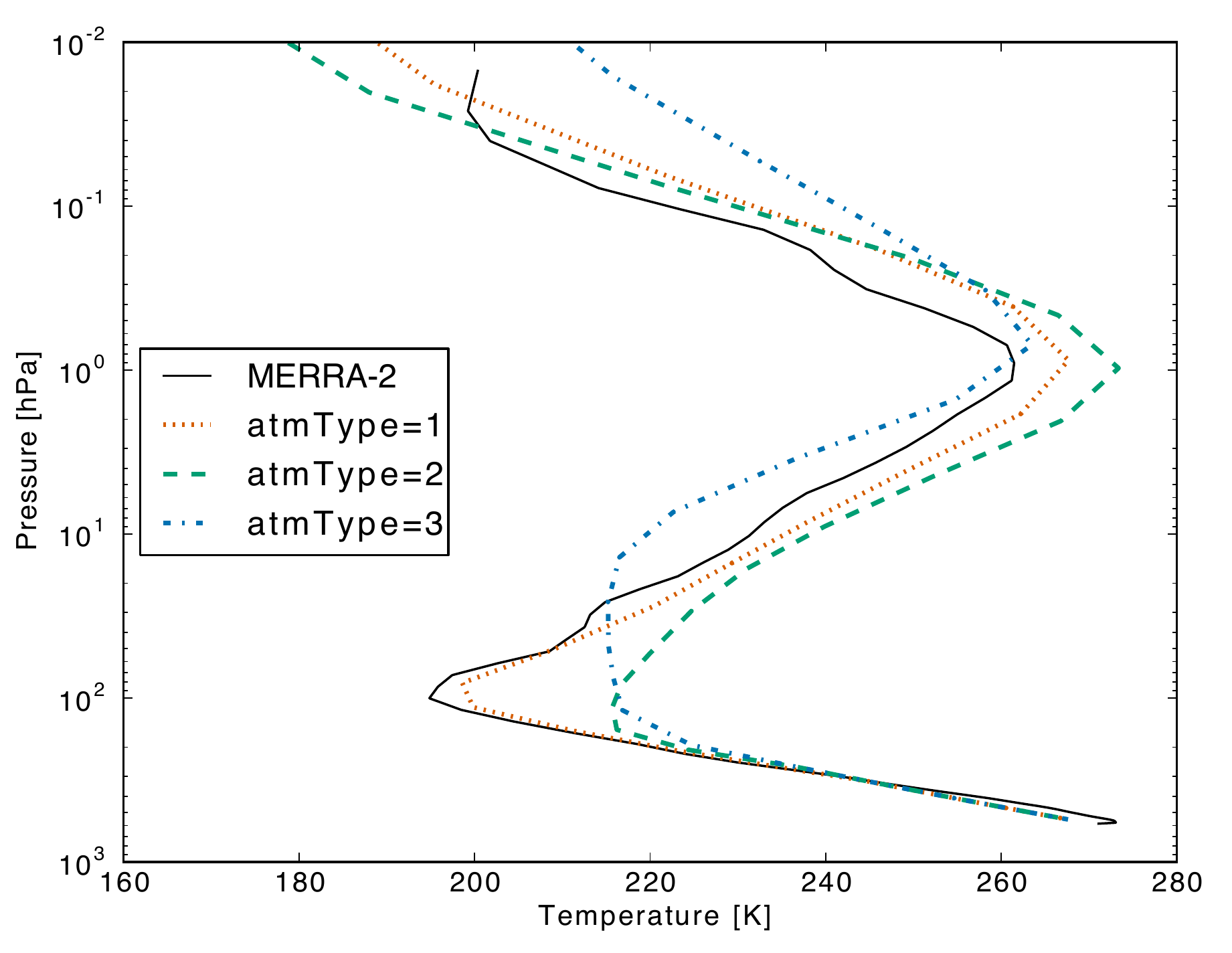}{0.5\textwidth}{(b)}}
\gridline{\leftfig{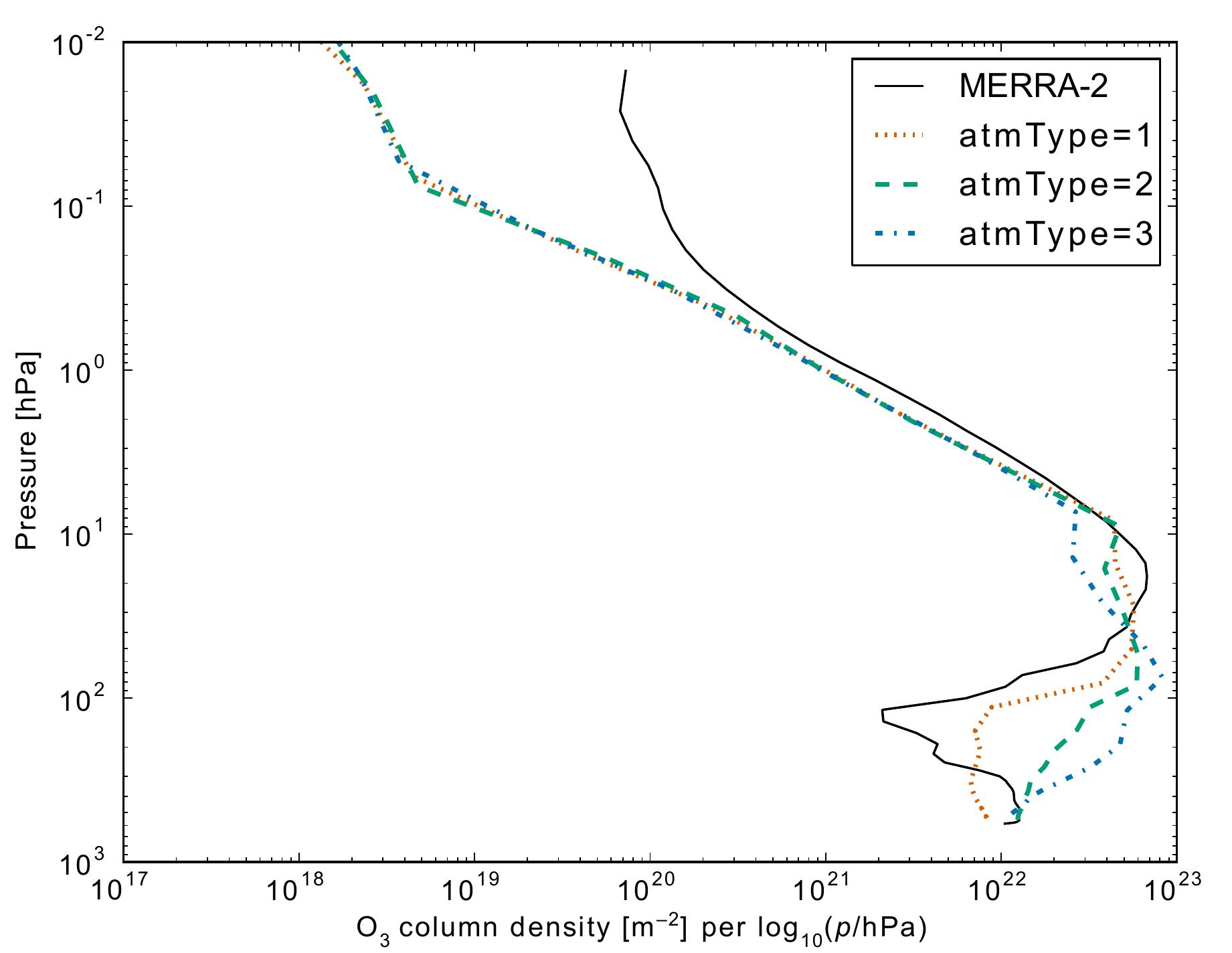}{0.5\textwidth}{(c)}
          \rightfig{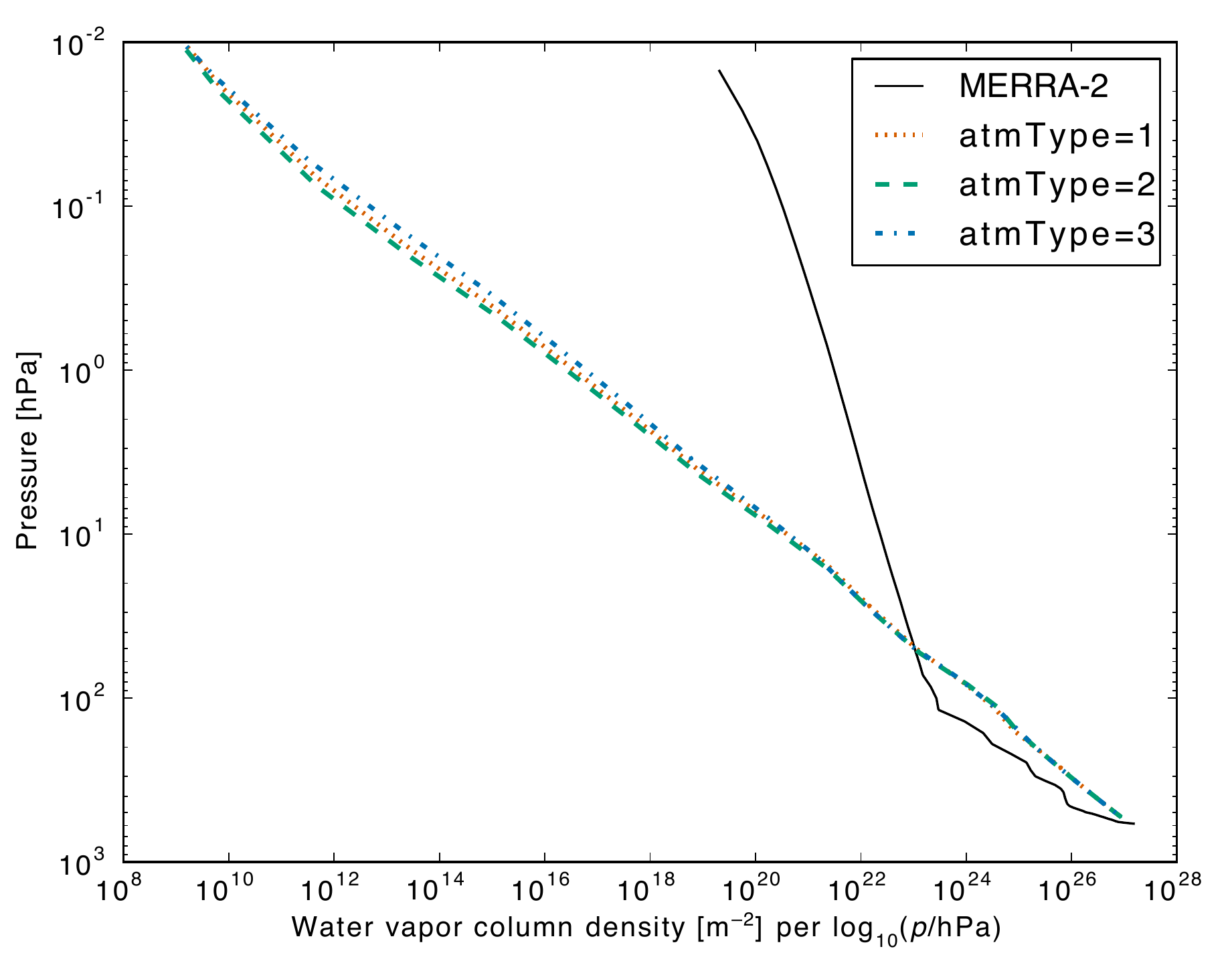}{0.5\textwidth}{(d)}}
\caption{(a) The pressure $p$ as a function of altitude for
uid://A002/Xb020f7/X45ad.
The three-hourly instantaneous datasets from MERRA-2
\citep{merra2asm,merra2chm} at the closest
spatial and temporal grids to those of the ALMA dataset,
namely ($67\fdg 5$ W, $23\fdg 0$ S) and 2016-03-05 09:00 UTC,
respectively, are compared with the ATM models with
$\mathtt{atmType}=1$ ({\it tropical\/}), 2 ({\it mid latitude summer\/}),
and 3 ({\it mid latitude winter\/}).
(b) The vertical profile of the temperature.
(c) The vertical profile of the O$_3$ column density per unit
$\log_{10}(p/\mathrm{hPa})$.
(d) Same as (c), but for water vapor.
\label{fig:merra2}}
\end{figure}

The residual line features after correction may be due to
discrepancies between the ATM models and the actual atmospheric profiles.
We evaluate the discrepancy by comparing the ATM model
for uid://A002/Xb020f7/X45ad (Figure \ref{fig:otherbands}b)
with meteorological reanalysis datasets from the second Modern-Era
Retrospective analysis for Research and Applications
\citep[MERRA-2;][]{2017JCli...30.5419G}.
Figure \ref{fig:merra2} shows the vertical profiles of the pressure ($p$),
temperature, and the column densities of O$_3$ and H$_2$O molecules.
The ${\tt atmType}=1$ profiles of temperature and O$_3$ column density
are closest among the three models to those from MERRA-2.
This appears to be consistent with the fact that the correction works
best with ${\tt atmType}=1$.
The O$_3$ profiles, though, reveal a few differences between the
MERRA-2 data and the ATM models:
\begin{enumerate}
\item Excess O$_3$ abundance is seen in the MERRA-2 data at the mesosphere
($p\lesssim 0.3\;\mathrm{hPa}$) compared with the ATM models.
This is attributed to the diurnal variation of mesospheric ozone abundance
\citep[e.g.,][]{1982Natur.296..133V,1998JGR...103.6189P};
the excess disappears in the
daytime MERRA-2 data, although not shown here.
\item The O$_3$ abundance in the ATM models at the lower mesosphere and
upper stratosphere ($\approx 0.3$--5 hPa) is
slightly less than that in the MERRA-2 data.
\item At the middle of the stratosphere ($\approx 10$--30 hPa),
the ATM model ($\mathtt{atmType}=3$, in particular) provides less
O$_3$ than MERRA-2.
\item The O$_3$ abundance in the ATM models is largely discrepant with
the MERRA-2 data at the lower stratosphere and troposphere
($\gtrsim 50\;\mathrm{hPa}$).
However, due to the large pressure broadening compared with the
instantaneous bandwidth, the impact of this component
on the application of the method to ALMA TP Array data is not
significant (i.e., mostly eliminated by spectral baseline subtraction).
\end{enumerate}

It is expected that the correction may improve if the realistic
atmospheric profiles from MERRA-2 can be used.
However we cannot quantitatively prove this expectation, because CASA's
{\tt atmosphere} tool does not provide the interface to directly control
the profiles (except for the temperature, cf.\ Table \ref{tab:parameters}).
Instead, we correlate the discrepancies described above with the
residual line features in a semi-quantitative way.

\begin{figure}
\gridline{\leftfig{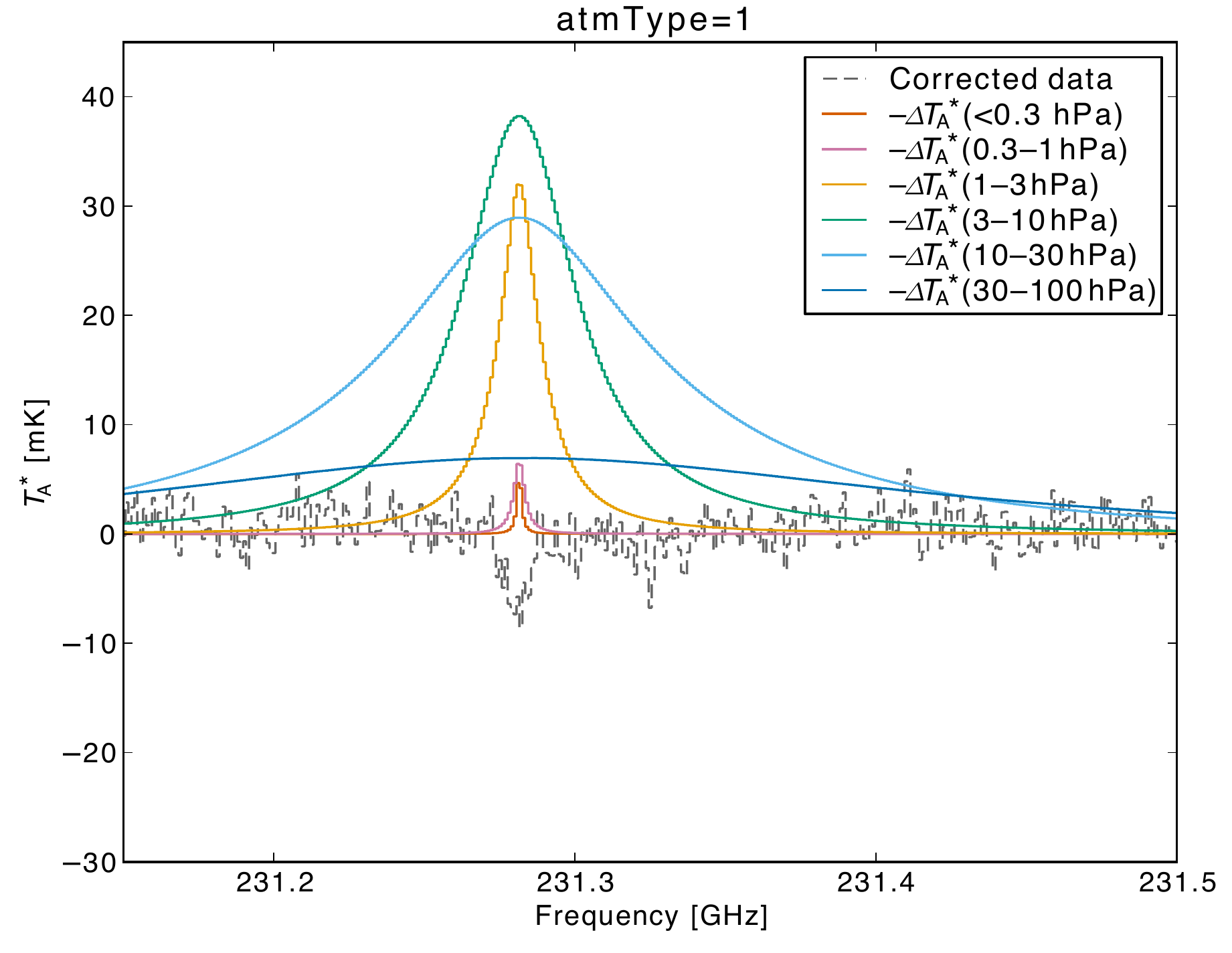}{0.5\textwidth}{(a)}
          \rightfig{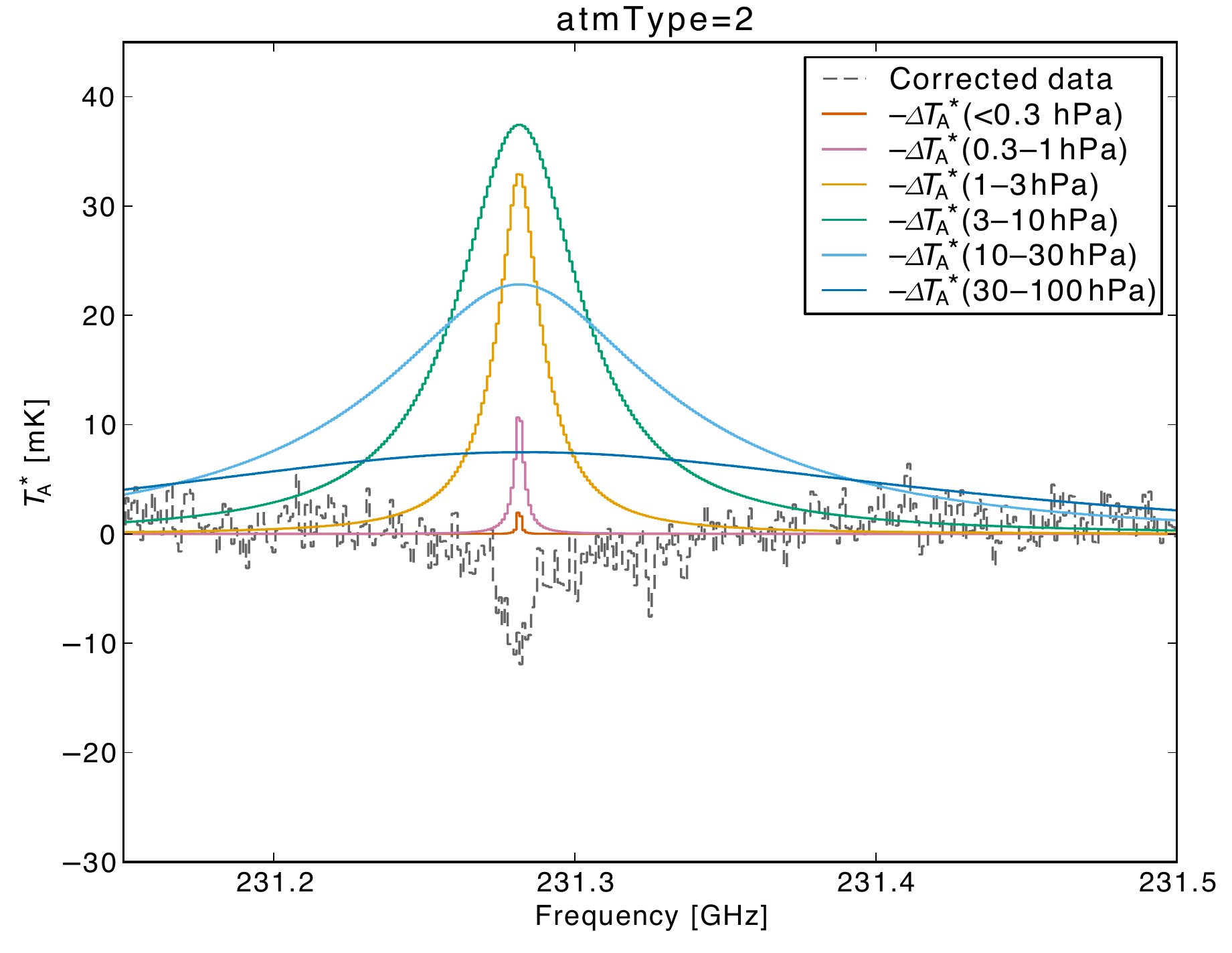}{0.5\textwidth}{(b)}}
\gridline{\leftfig{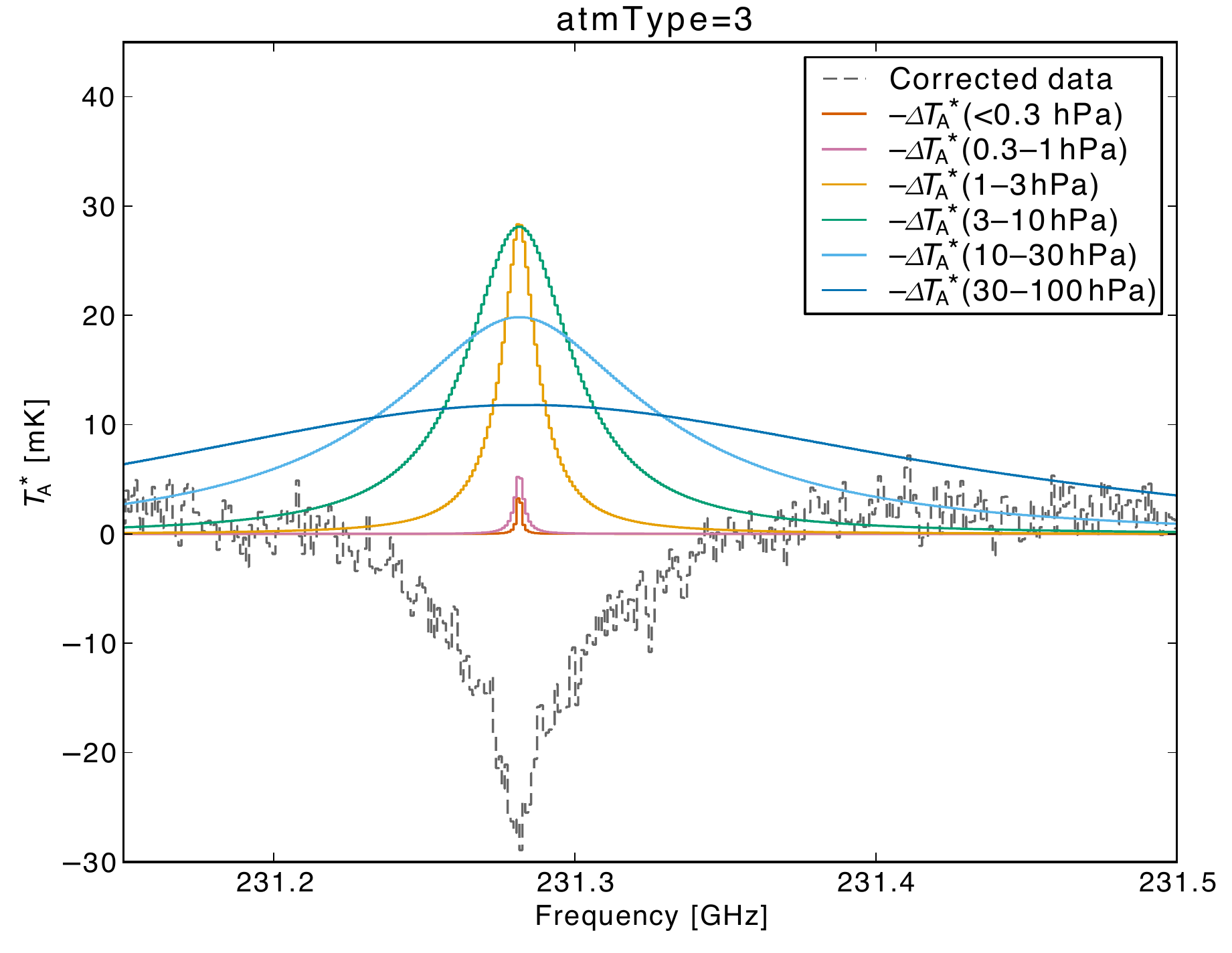}{0.5\textwidth}{(c)}
          \rightfig{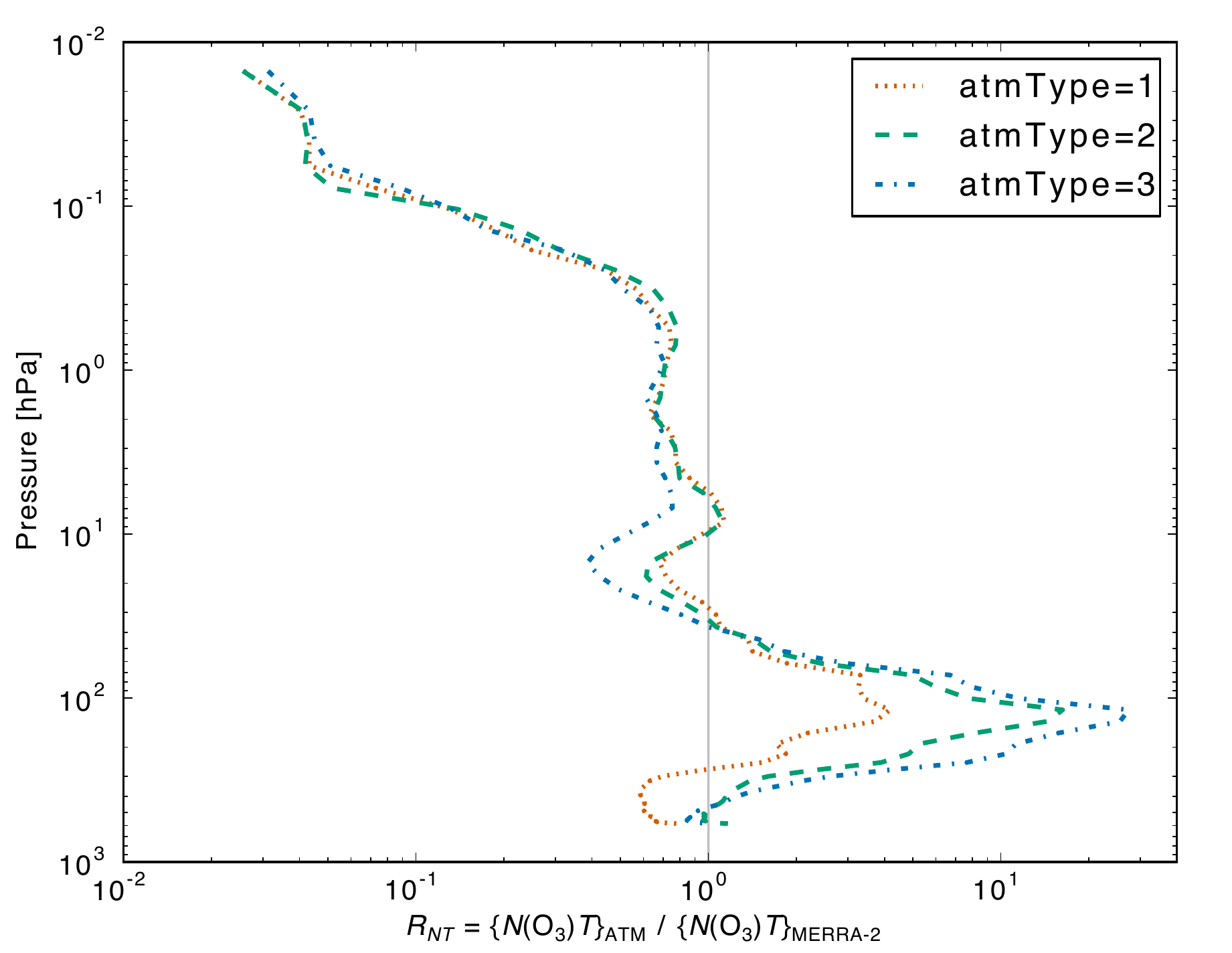}{0.5\textwidth}{(d)}}
\caption{
(a), (b), (c) The ${\mit\Delta}\tastar$ components corresponding to
six layers of the atmosphere ($p<0.3$, 0.3--1, 1--3, 3--10,
10--30, and 30--100 hPa) and the corrected spectra using
$\mathtt{atmType}=1$, 2, and 3, respectively, for
uid://A002/Xb020f7/X45ad.
The signs of the ${\mit\Delta}\tastar$ spectra are flipped
for visibility.
Note that the quasi-continuum
component is already removed by spectral baseline subtraction,
which is done at $\simeq 500$ MHz from the line center.
(d) 
The ATM-to-MERRA-2 ratio of the product of the
O$_3$ column density and temperature, $R_{NT}$.
\label{fig:dtaperaltitude}}
\end{figure}

We estimate the ${\mit\Delta}\tastar(1)$ (hereafter referred to as
${\mit\Delta}\tastar$ for simplicity) component for a layer of
the atmosphere within a certain range of pressure (altitude) by
taking the difference of ${\mit\Delta}\tastar$ spectra
calculated using two values of {\tt maxAltitude}.
Figure \ref{fig:dtaperaltitude}a--c show the ${\mit\Delta}\tastar$
components corresponding to six layers ($p<0.3$, 0.3--1, 1--3, 3--10,
10--30, and 30--100 hPa) along with the corrected spectra with
$\mathtt{atmType}=1$, 2, and 3 (same as those shown in Figure
\ref{fig:otherbands}b), respectively, for EB uid://A002/Xb020f7/X45ad.
Figure \ref{fig:dtaperaltitude}d shows the ratio (hereafter $R_{NT}$),
between the ATM model and MERRA-2 data,
of the product of the O$_3$ column density (Figure \ref{fig:merra2}c)
and temperature (Figure \ref{fig:merra2}b).
Since the 231.28 GHz O$_3$ line is optically thin (cf.\ Figure
\ref{fig:transmission}, unless the airmass is very high),
the product of the column density and temperature is roughly
proportional to the observed atmospheric line emission
originating in the corresponding altitude.

Now we revisit the discrepancies between the MERRA-2 data and
the ATM model described above:
\begin{enumerate}
\item Under-abundance of O$_3$ in the ATM models at
$p\lesssim 0.3\;\mathrm{hPa}$.
The $R_{NT}$ is $\approx 0.5$ at 0.3 hPa and gets lower
at higher altitude.
Therefore the corresponding ${\mit\Delta}\tastar$ component is
under-corrected by a factor of two or more.
However, since the amplitude of this component is not very high
(comparable to the noise level), its impact on the correction
is likely not significant for this particular dataset.
\item Under-abundance at $\approx 0.3$--5 hPa.
The corresponding ${\mit\Delta}\tastar$ component is estimated,
as the sum of the 0.3--1 and 1--3 hPa spectra in Figure
\ref{fig:dtaperaltitude}, to have the peak intensity of
$\simeq 40$ mK and the full width at half maximum (FWHM) of
$\simeq 10$ MHz.
Since $R_{NT}$ is typically 0.7--0.8, this component is
under-corrected by $\mathrm{a~few}\times 10\%$, i.e.,
$\simeq 10$ mK.
This seems to largely explain the residual feature around the
line center, albeit possible over-correction.
\item Under-abundance (in particular ${\tt atmType}=3$) at
$\approx 10$--30 hPa.
The corresponding ${\mit\Delta}\tastar$ component has
$\simeq 20$ mK peak intensity and $\simeq 100$ MHz FWHM.
The $R_{NT}$ varies with altitude and ranges between $\simeq 0.4$--1.
As a rough estimate, we assume the typical ratio to be 0.7,
i.e., the component is under-corrected by about
$40\% \simeq 8$ mK.
It, along with the under-abundance at higher altitude
(typically $R_{NT}\simeq 0.7$), is likely responsible to the
residual feature with wings.
\end{enumerate}

We showed that the correction may become better
if the atmospheric model is improved.
However, such an improvement is not available in the short term.
Therefore, in the following subsection,
we seek best-fitting parameters for the current model.

\subsection{Seasonal and Daily Variations}\label{sec:season}

For the datasets shown in Figure \ref{fig:otherbands}, the
{\it tropical\/} model ($\mathtt{atmType}=1$) seems to work the best.
However, the best parameter(s) may vary from season to season, and/or
between day and night.
In order to see if such a variation exists, we analyze the 231.28 GHz
O$_3$ line in the sixteen datasets listed in Table \ref{tab:seasons}.
The line is chosen because it has often been observed simultaneously
with the CO $J=2\mbox{--}1$ line at 230.54 GHz.
The datasets are arbitrarily picked from the public data in the archive,
spreading through the seasons, daytime (between midday and sunset)
and nighttime (between midnight and sunrise).
In addition to {\tt atmType}, we also change secondary parameters,
namely the lapse rate ({\tt dTem\_dh}) and water scale height ({\tt h0}).
The parameter set $(\mathtt{dTem\_dh}\;[\mathrm{K\,km^{-1}}],
\mathtt{h0}\;[\mathrm{km}]) = (-6.8, 1.5)$ estimated by
\citet{ALMAMemo496}, as well as the function's default $(-5.6, 2.0)$,
is attempted.
Upon the comparison, the data are smoothed/resampled to the common
resolution/spacing of 2 MHz.

\begin{deluxetable*}{llllrrrrrl}
\tablecolumns{9}
\tablewidth{0pc}
\tablecaption{The properties of the EBs used to study the seasonal/daily variation\label{tab:seasons}}
\tablehead{
  \colhead{EB UID\tablenotemark{a}} & \colhead{Project code} & \colhead{SB name\tablenotemark{b}} & \colhead{Date and time\tablenotemark{c}} & \colhead{PWV} & \colhead{Temperature} & \colhead{Pressure} & \colhead{Humidity} & \colhead{In Fig.\ \ref{fig:seasons}}\\
  \colhead{} & \colhead{} & \colhead{} & \colhead{UTC} & \colhead{(mm)} & \colhead{(K)} & \colhead{(hPa)} & \colhead{(\%)} & \colhead{}
}
\startdata
Xc92012/X16e1\tablenotemark{d} & 2016.1.01346.S & AGAL010.\_a & 2018-01-14 17:03 & 1.63 & 275.4 & 554.9 & 37 & Yes\\
Xb046c2/X406f & 2015.1.00121.S & M83\_a & 2016-03-09 05:05 & 1.27 & 268.4 & 556.6 & 60 & No\\
Xb12f3b/Xafe8 & 2015.1.00121.S & M83\_b & 2016-04-01 06:39 & 1.43 & 272.9 & 558.3 & 11 & Yes\\
Xc04da7/X7f4 & 2015.1.00121.S & M83\_a & 2017-05-14 05:12 & 1.07 & 264.7 & 556.6 & 51 & No\\
Xb499c3/X108ef & 2015.1.00357.S & G286\_5\_a & 2016-06-23 19:48 & 0.85 & 272.1 & 554.8 & 10 & Yes\\
Xcf196f/X26f9 & 2017.1.00716.S & G333.01\_a & 2018-06-28 04:33 & 0.72 & 263.6 & 553.6 & 56 & Yes\\
Xb54d65/X37e0 & 2015.1.01539.S & G28.23\_a & 2016-07-13 06:54 & 0.54 & 261.7 & 553.3 & 16 & No\\
Xb57bb5/X2839 & 2015.1.00357.S & G286\_5\_a & 2016-07-17 18:07 & 0.52 & 270.4 & 554.9 & 3 & No\\
Xb638bc/X1df2 & 2015.1.00667.S & AG22.36+\_l & 2016-08-03 05:02 & 0.72 & 264.6 & 556.5 & 14 & No\\
Xb66ea7/X599c & 2015.1.00357.S & G286\_5\_a & 2016-08-09 17:33 & 0.31 & 270.3 & 554.1 & 4 & No\\
Xd1daeb/X42d0 & 2017.1.00093.S & YSO45\_a & 2018-09-11 05:54 & 0.64 & 267.6 & 555.5 & 13 & No\\
Xb9356a/X3398 & 2016.1.00203.S & LMC\_N166\_a & 2016-10-10 07:19 & 0.68 & 264.2 & 553.8 & 36 & Yes\\
Xb903d6/X324e & 2016.1.00386.S & M83\_c & 2016-10-04 18:36 & 0.70 & 273.1 & 554.9 & 5 & Yes\\
Xba1cd8/X41d4 & 2016.1.00386.S & M83\_a & 2016-11-03 16:20 & 0.65 & 277.3 & 556.7 & 1 & No\\
Xd704f8/X74a5\tablenotemark{d} & 2018.1.00443.S & G343.756\_a & 2018-12-22 15:40 & 2.04 & 275.0 & 556.0 & 27 & No\\
Xd704f8/Xd618 & 2018.1.00770.S & Hummingb\_a & 2018-12-23 05:30 & 0.61 & 269.4 & 555.7 & 26 & Yes
\enddata
\tablenotetext{a}{The common prefix, ``uid://A002/'', is omitted.}
\tablenotetext{b}{The common suffix, ``\_06\_TP'', is omitted.}
\tablenotetext{c}{The centroid of the on-source integrations.
  The Chilean standard time and summer time are $-4$:00 and $-3$:00,
  respectively, from UTC.}
\tablenotetext{d}{Multiple targets were observed in the EB.
  Only the first one is used for the analysis.}
\end{deluxetable*}%

\begin{figure}
\gridline{\leftfig{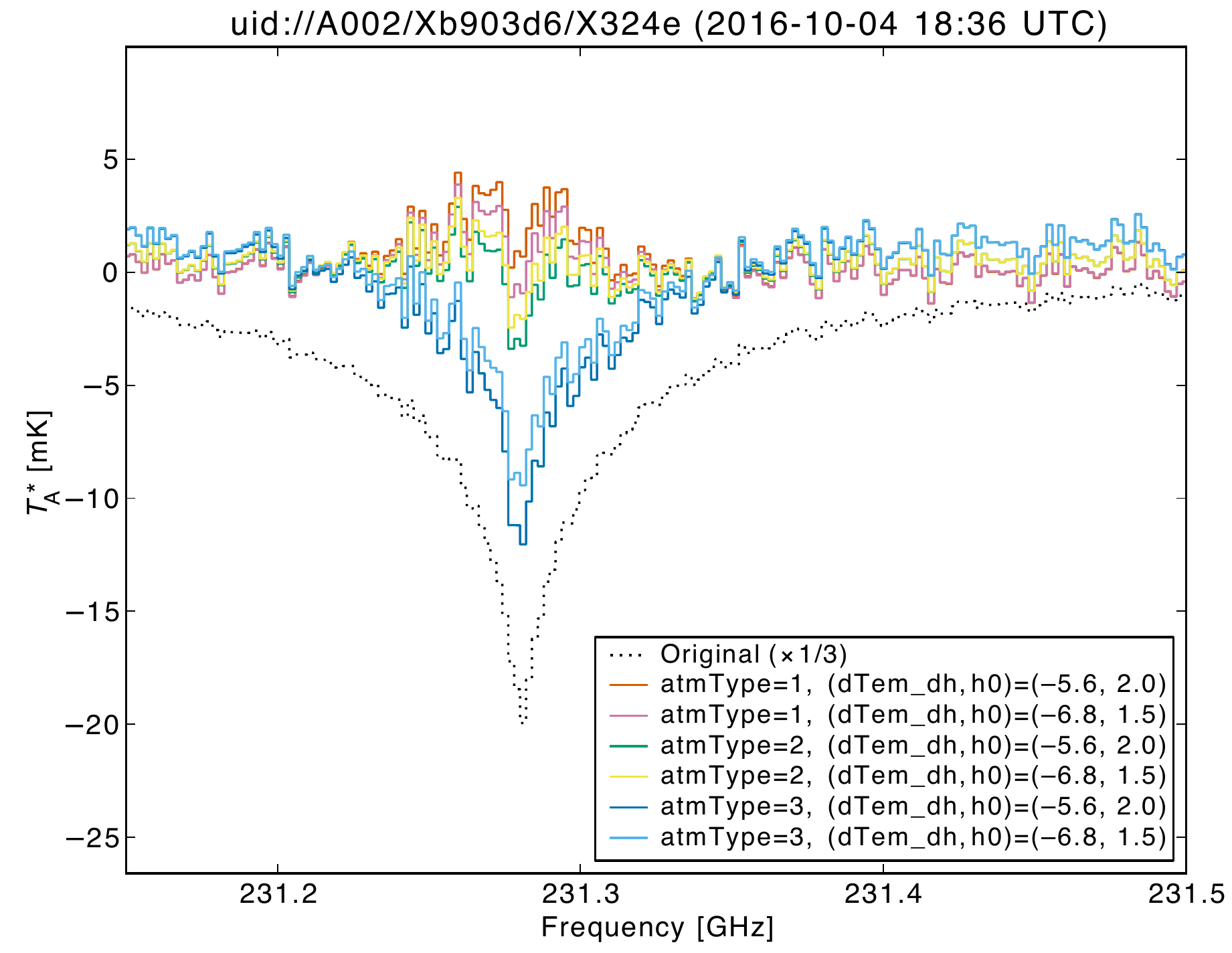}{0.34\textwidth}{}
          \rightfig{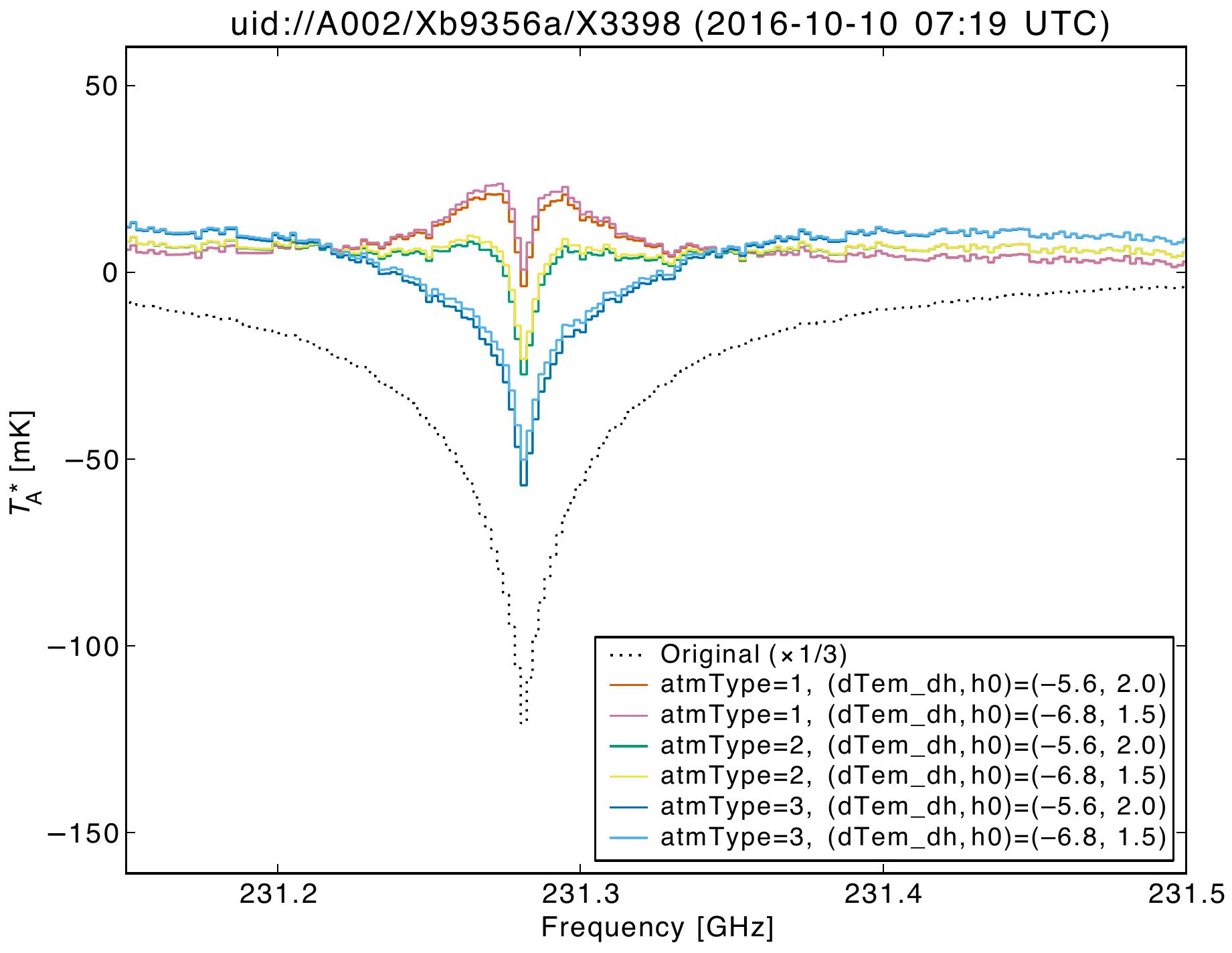}{0.34\textwidth}{}}
\gridline{\leftfig{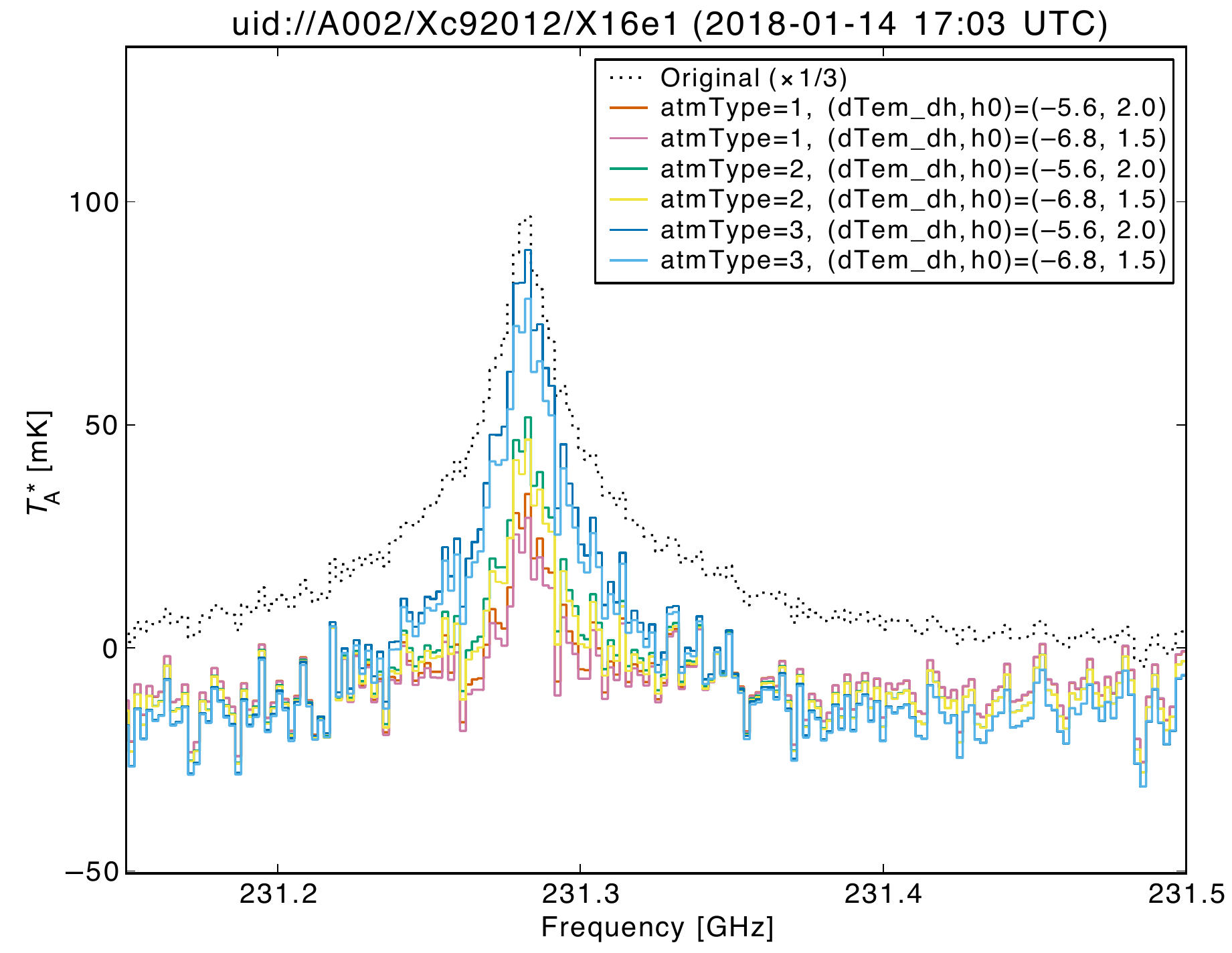}{0.34\textwidth}{}
          \rightfig{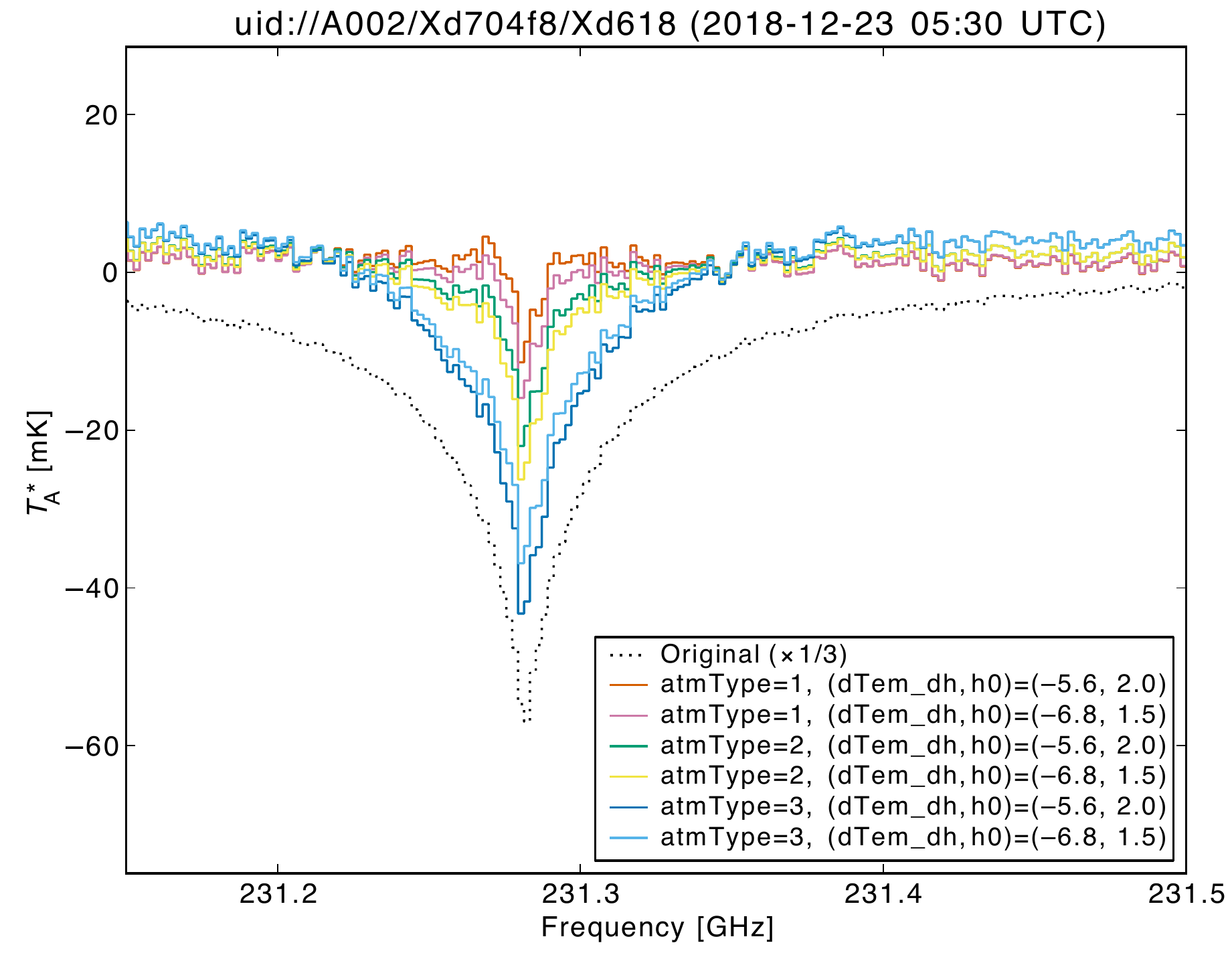}{0.34\textwidth}{}}
\gridline{\rightfig{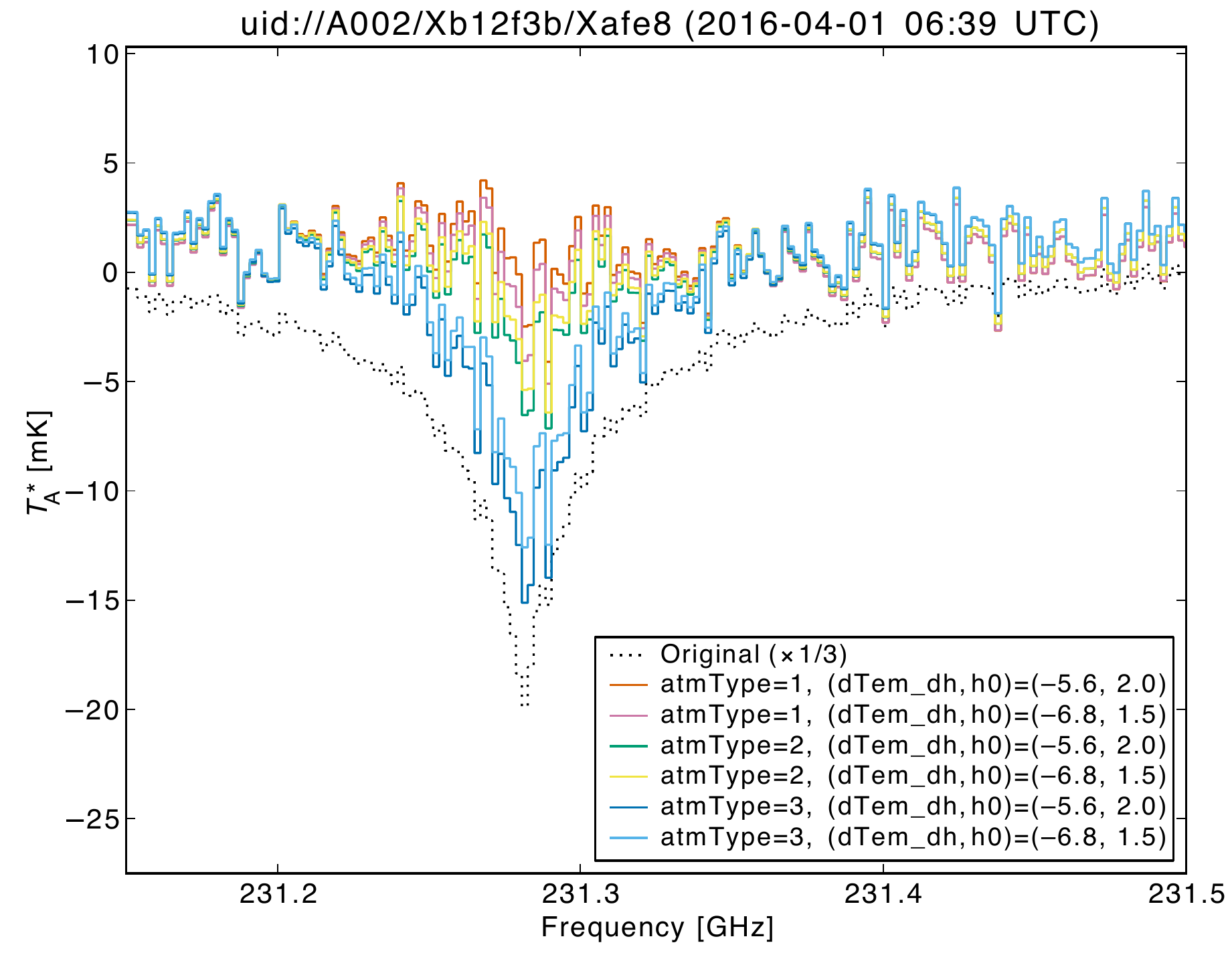}{0.34\textwidth}{}}
\gridline{\leftfig{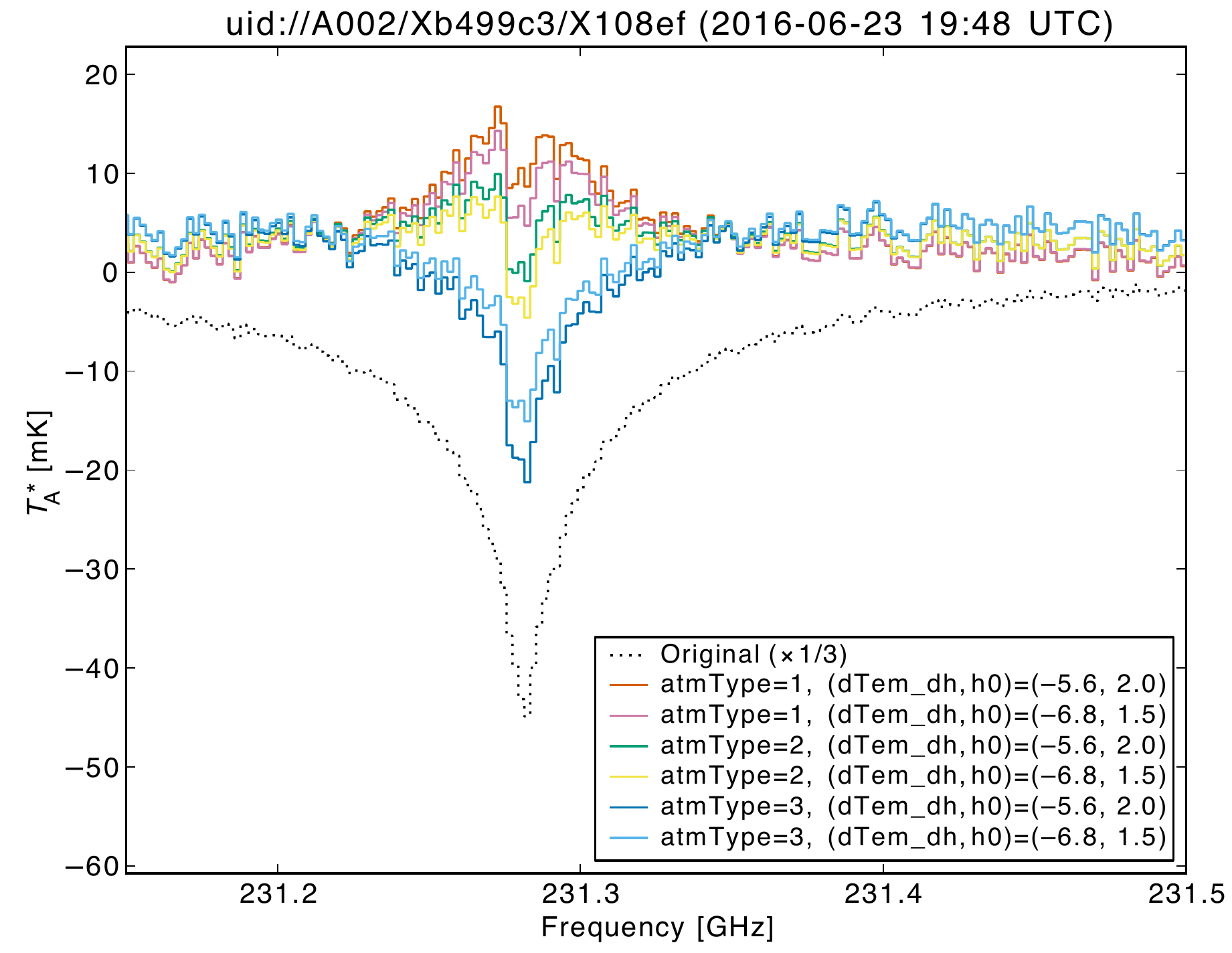}{0.34\textwidth}{}
          \rightfig{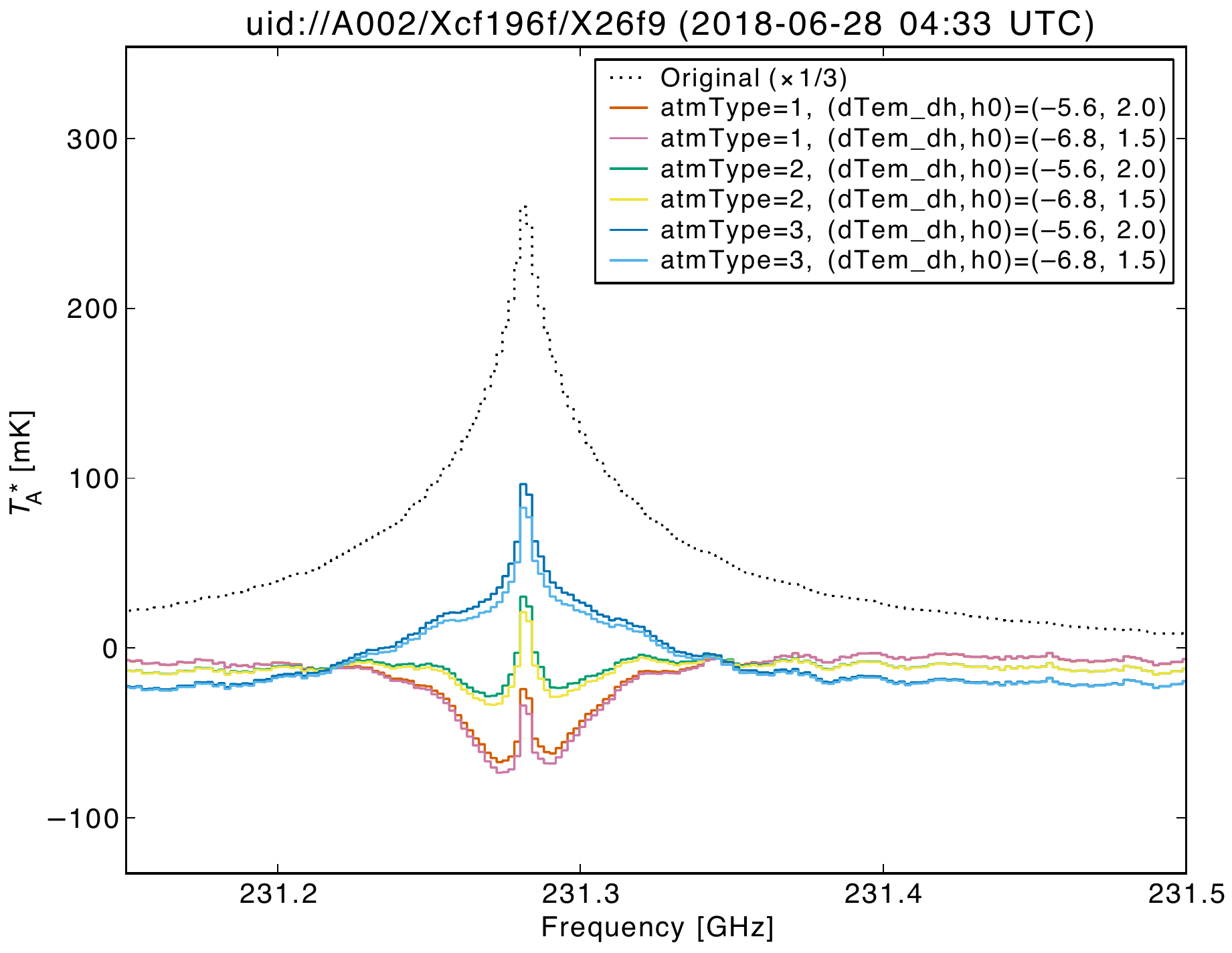}{0.34\textwidth}{}}
\caption{The uncorrected and corrected spectra of the 231.28 GHz O$_3$
line from the seven EBs.
The EB UID and observing date/time are shown at the top of each panel.
The panels are sorted by the season (spring, summer, autumn, and winter
from top to bottom), and those on the left and right sides are for
the data observed in the daytime and nighttime, respectively.
The uncorrected spectra (dotted lines) are divided by 3 so that they
fit into the axes scale.
\label{fig:seasons}}
\end{figure}

The spectra for the representative seven datasets (observed around
the solstices and equinoxes, daytime and nighttime) are shown in
Figure \ref{fig:seasons}.
For the data taken in summer (the second row), the residual line is
minimized by using the {\it tropical\/} model ($\mathtt{atmType}=1$).
For those taken in winter (the fourth row), on the other hand,
the {\it mid latitude summer\/} model ($\mathtt{atmType}=2$) works better,
the {\it tropical\/} model over-correcting the inner line wings.
Varying the parameter set $(\mathtt{dTem\_dh}, \mathtt{h0})$ causes
much smaller difference than varying {\tt atmType} does and
it is hard to tell which one is better.

\begin{figure}
\gridline{\fig{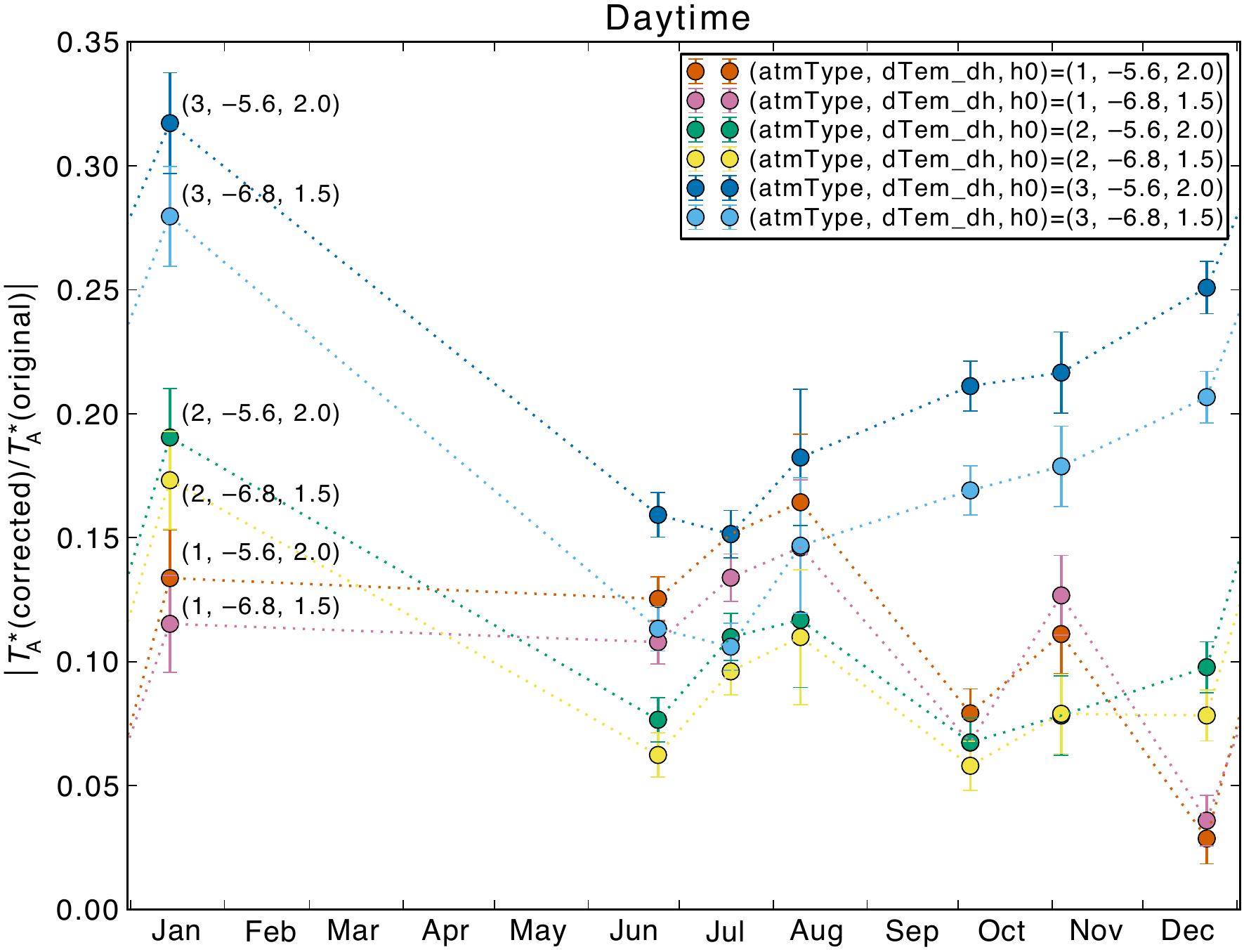}{0.7\textwidth}{(a)}}
\gridline{\fig{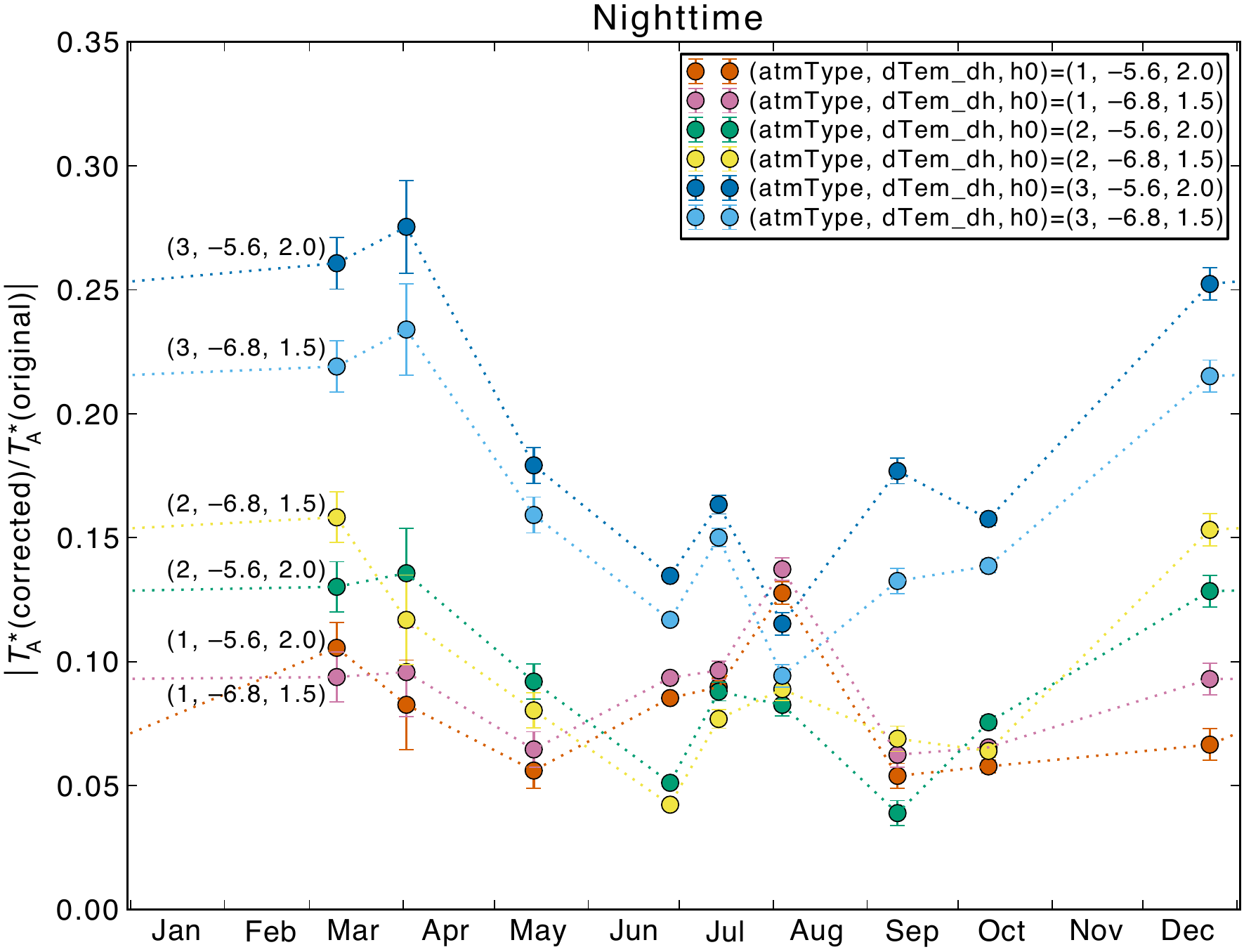}{0.7\textwidth}{(b)}}
\caption{The ratios between the maximum intensities of the
corrected and original 231.28 GHz O$_3$ line spectra
for the datasets listed in Table \ref{tab:seasons},
as a function of observing date.
(a) The datasets observed in the daytime, (b) those in the nighttime.
The error bars indicate the $1\sigma$ noise.
Note that the maximum intensities in the corrected spectra
do not necessarily appear at the center of the line
(see Figure \ref{fig:seasons}).
\label{fig:suppressionratio}}
\end{figure}

The results from the sixteen datasets are shown in Figure
\ref{fig:suppressionratio};
the ratio between the maximum line intensities of the
corrected and original spectra is plotted as a function of
observing season, for the daytime and nighttime observations.
These comparisons indicate that, as already demonstrated for the
representative seven datasets, the {\it tropical\/}
($\mathtt{atmType}=1$) and {\it mid latitude summer\/} (2) models work
best for the data taken in summer and winter, respectively,
at least for this particular line.
The maximum line intensities are suppressed by factors of 
$\approx 10$--20, if the appropriate parameters are chosen.
The difference between the daytime and nighttime observations is not
clear.  Although the diurnal variation of mesospheric O$_3$ abundance is
known to exist, the amplitude of the emission from that altitude
is low (Section \ref{sec:parameters}).

\subsection{Molecules Other Than Ozone}

The narrow lines that show up in the mm to sub-mm regime
(Figure \ref{fig:transmission}) are mostly from O$_3$ molecules.
The line profiles of other molecules that are abundant in the lower
atmosphere (e.g., H$_2$O and O$_2$) are highly pressure-broadened
and hence outside the scope of the analyses in this section
(i.e., the lines are much broader than the ALMA's instantaneous
bandwidth, 2 GHz per baseband).

\begin{figure}
\plotone{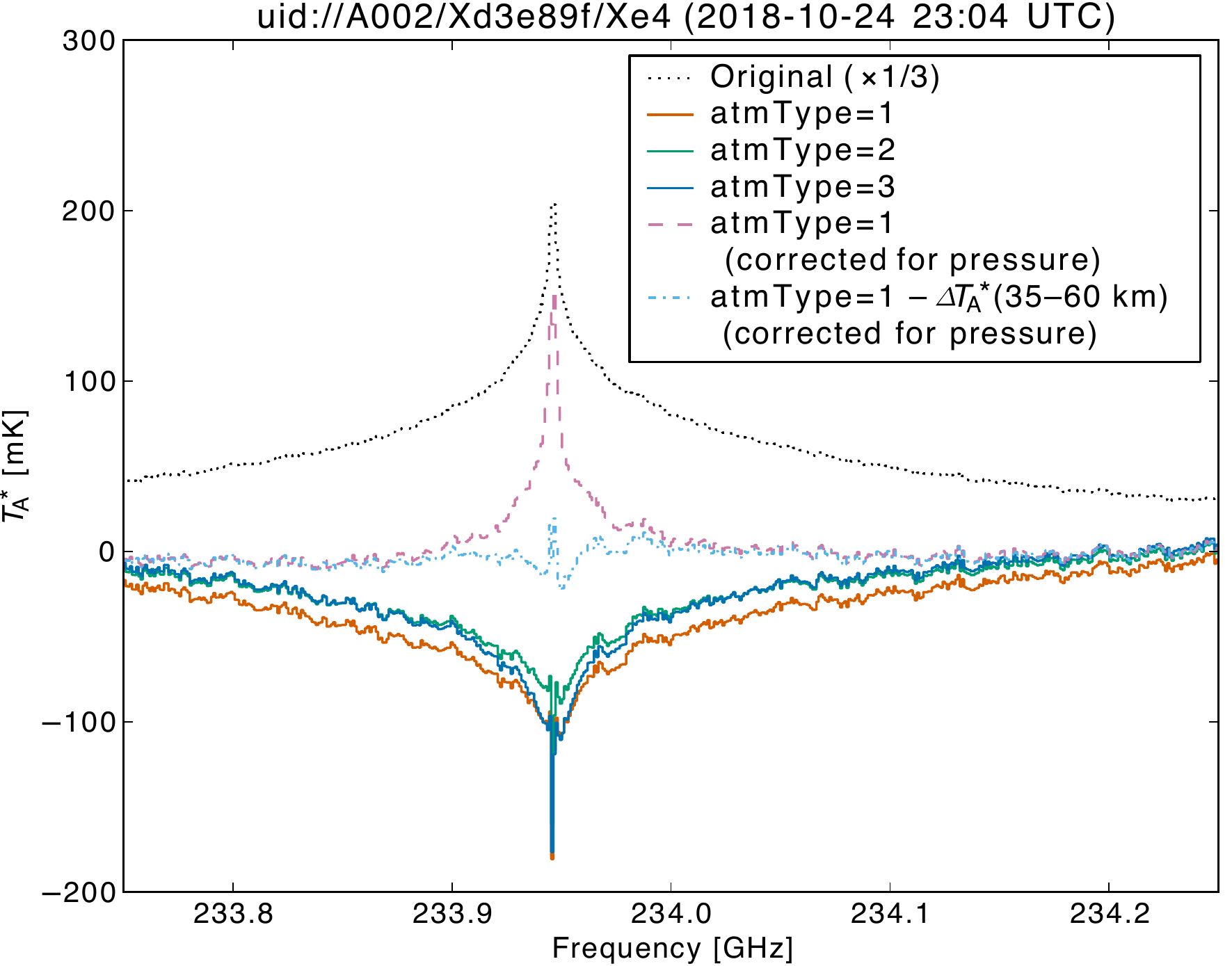}
\caption{The uncorrected and corrected average $\tastar$ spectra of
the 233.95 GHz $^{16}$O$^{18}$O line from the EB
uid://A002/Xd3e89f/Xe4 (SB 24013+04\_a\_06\_TP, project 2018.1.00443.S).
The uncorrected spectrum (dotted line) is divided by 3.
The corrected spectra using ${\tt atmType}=1$, 2, and 3 are shown
as solid lines.
For ${\tt atmType}=1$, the spectrum corrected for the pressure
is shown as a dashed line, and another spectrum from which
the pressure-corrected ${\mit\Delta}\tastar$ for 35--60 km altitude
is further subtracted is displayed as a dash-dotted line
 (see the text).
Note that the small wiggles seen in the line wings are unrelated to
the correction.
\label{fig:O18O}}
\end{figure}

Figure \ref{fig:O18O} shows an exceptional case; the 233.95 GHz
line of the oxygen isotopologue $^{16}$O$^{18}$O before and after
the correction.
Due to the species' rarity, the narrow peak of the line originating
from the upper atmosphere is visible \citep[cf.][]{1995JQSRT..54..931P}.
The correction is not as good as that for the O$_3$ lines presented
earlier; all the three {\tt atmType} values similarly result in
over-correcting the line, with a sharp feature at the line center.

\begin{figure}
\plotone{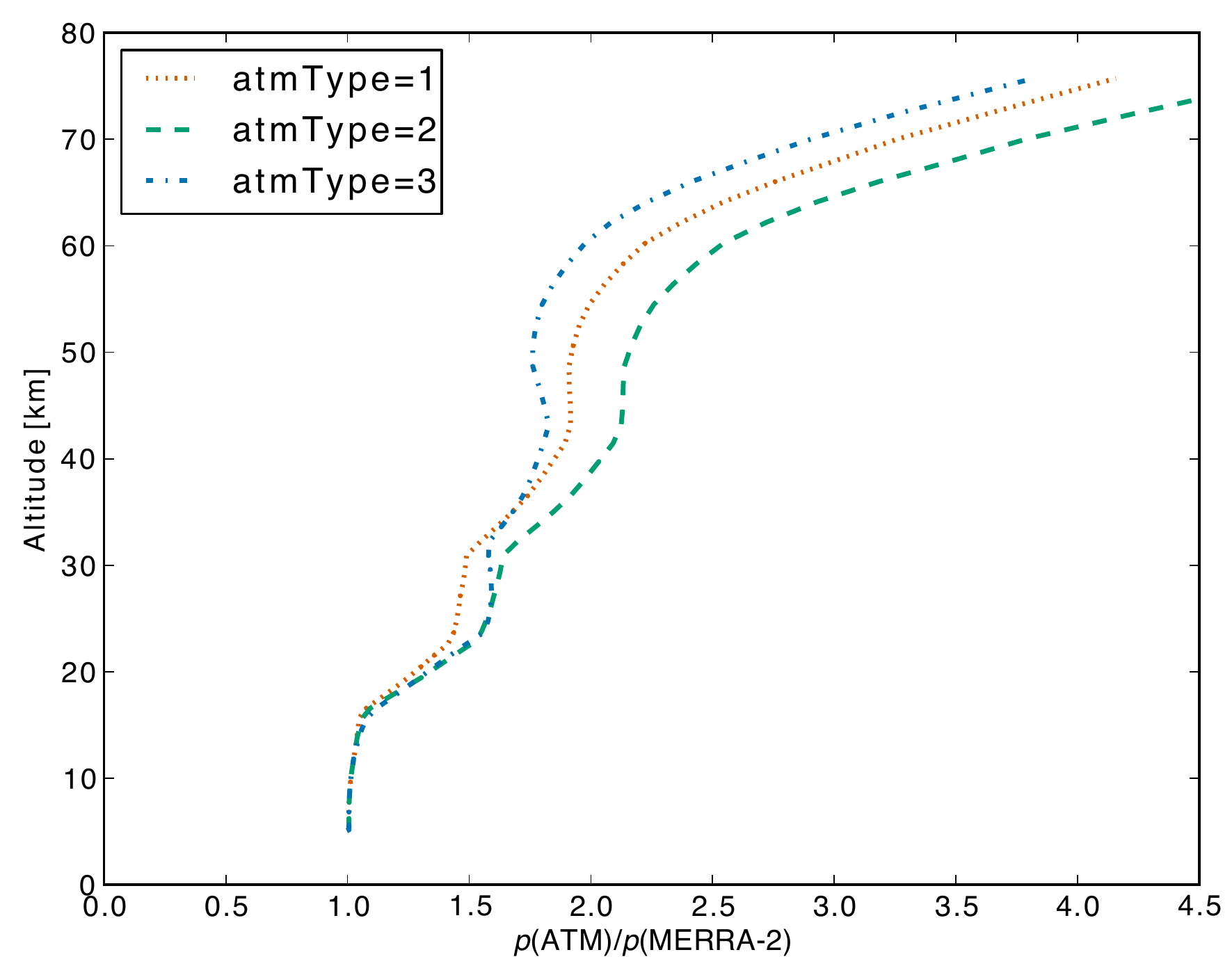}
\caption{The ratio between the vertical pressure profiles from ATM
and MERRA-2 for uid://A002/Xd3e89f/Xe4.
The MERRA-2 three-hourly instantaneous dataset \citep{merra2asm}
at ($67\fdg 5$ W, $23\fdg 0$ S), 2018-10-25 00:00 UTC is used.
\label{fig:Pratio}}
\end{figure}

The poor correction can be, at least partially, attributed to the
difference between the vertical atmospheric profiles
from ATM and MERRA-2.
The emission from a given altitude is proportional to the
product of the number density of the molecule and temperature
as mentioned in Section \ref{sec:parameters} and, in the case of
oxygen (which is supposed to be well-mixed in the atmosphere),
it is also proportional to the pressure (assuming the ideal gas law).
Figure \ref{fig:Pratio} shows the pressure ratio between ATM and
MERRA-2 as a function of altitude.
The typical ratios at the altitudes of 20--35, 35--60, and $>60$ km
are $\simeq 1.5$, 2, and 3, respectively, with $\mathtt{atmType}=1$.
That is, the correction ${\mit\Delta}\tastar$ is likely
overestimated by these factors at these altitude ranges.
Therefore we determine the ${\mit\Delta}\tastar$ components for
20--35, 35--60, and $>60$ km altitudes as we did in Section
\ref{sec:parameters}, and divide them by 1.5, 2, and 3, respectively.

The result is shown in Figure \ref{fig:O18O} as a dashed line.
Although the correction is still not very accurate near the
center of the line (nevertheless, the peak intensity is suppressed
by about a factor of 4 relative to its uncorrected intensity),
the line wing is much better corrected.
Empirically, the residual can be mostly eliminated by further
subtracting the pressure-corrected ${\mit\Delta}\tastar$ for
the altitude range of 35--60 km (in other words, if the pressure
correction by a factor of 2 was not applied), as shown in Figure
\ref{fig:O18O} as a dash-dotted line.
Although a naive interpretation of this calculation may lead to
an incorrect estimate of the abundance by a factor of 2,
this interpretation seems unlikely because
oxygen is expected to be well-mixed in the atmosphere.
We leave the root cause of the less-than-optimal modeling unknown;
it is beyond the scope of this article to completely diagnose
the limitations of the built-in ATM atmosphere models
and comparison with MERRA-2.

\subsection{Caveats}

We have analyzed the seasonal/daily variation of the 231.28 GHz O$_3$
line in Section \ref{sec:season} and demonstrated that the
{\it tropical\/} and {\it mid latitude summer\/} models work the best
for the data taken in summer and winter, respectively.
This might possibly be explained qualitatively as follows;
the atmosphere at the ALMA site (located near the tropic of Capricorn,
the zenith angle of the sun at the transit ranging from $90\arcdeg$
in summer to $44\arcdeg$ in winter) is tropical-like in summer and
mid-latitude-like in winter.
However, this could also be a coincidence.
In fact, the 481.62 GHz O$_3$ line data shown in Figure
\ref{fig:otherbands}d is contradictory, the {\it tropical\/} model being
the best while the dataset was taken in August (winter).
Therefore a future study using a larger set of data is required.
Such a study might also contribute to improving the atmospheric model.

In our implementation, the differences of the $\tsky$ and $\tau$ between
the ON and OFF positions are assumed to solely depend on elevation,
i.e., a stable and isotropic atmosphere is assumed.
In reality temporal and directional variation of the atmosphere causes
additional artifacts.
In the case of ALMA TP Array, theoretically, the short-timescale
variation of (the wet component of) the atmosphere can be traced
to some extent by using the data from the water vapor radiometer
\citep{2013A&A...552A.104N}, which is equipped on each antenna and
records the 183 GHz water line emission at an interval of
$\approx 1\;\mathrm{s}$.
However, the possibility of improving the correction
by taking the short-timescale variation into account
has not been explored and will be a topic of a future study.

\section{Conclusions}

We described a method to mitigate the atmospheric artifacts in
single-dish radio spectra using the ATM model and presented the
results for the data taken with the ALMA TP Array.
The residual atmospheric line intensities were suppressed typically
by an order of magnitude, dependent upon the species and transitions.
The best-fitting value of one of the ATM model parameters,
{\tt atmType}, was demonstrated to be dependent on the observing seasons
for the 231.28 GHz O$_3$ line.
A further study of parameter optimization for other lines
may help better mitigate the residual lines in future data processing.
We also compared the ATM model with MERRA-2 meteorological
reanalysis datasets for two transitions; the 231.28 GHz O$_3$ and
233.95 GHz $^{16}$O$^{18}$O lines.
This comparison suggests that the extraction of these two transitions
from total power spectra is improved if the difference between the ATM
and MERRA-2 models is taken into account.
The method described in this article will be implemented as a task in
CASA version 6.2.
It will also be included in the pipeline data processing for the
ALMA TP Array in the future.
The expectation is that the use of the algorithm described in
this article will reduce the number of
observations that cause difficulties in the Quality Assurance Level 2
\citep[cf.][]{ALMATHCy8} due to atmospheric artifacts.

\acknowledgments

This article makes use of the following ALMA data:
ADS/JAO.ALMA\#2015.1.00121.S,
ADS/JAO.ALMA\#2015.1.00274.S,
ADS/JAO.ALMA\#2015.1.00357.S,
ADS/JAO.ALMA\#2015.1.00667.S,
ADS/JAO.ALMA\#2015.1.00717.S,
ADS/JAO.ALMA\#2015.1.00908.S,
ADS/JAO.ALMA\#2015.1.01539.S,
ADS/JAO.ALMA\#2016.1.00203.S,
ADS/JAO.ALMA\#2016.1.00386.S,
ADS/JAO.ALMA\#2016.1.01346.S,
ADS/JAO.ALMA\#2016.1.01541.S,
ADS/JAO.ALMA\#2017.1.00093.S,
ADS/JAO.ALMA\#2017.1.00716.S,
ADS/JAO.ALMA\#2018.1.00272.S,
ADS/JAO.ALMA\#2018.1.00443.S,
and
ADS/JAO.ALMA\#2018.1.00770.S.
ALMA is a partnership of ESO (representing its member states), NSF (USA)
and NINS (Japan), together with NRC (Canada), MOST and ASIAA (Taiwan),
and KASI (Republic of Korea), in cooperation with the Republic of Chile.
The Joint ALMA Observatory is operated by ESO, AUI/NRAO and NAOJ.
This article makes use of the MERRA-2 datasets
{\it inst3\_3d\_asm\_Nv} and {\it inst3\_3d\_chm\_Nv}
retrieved on 2020 Dec 9 via
Goddard Earth Sciences Data and Information Services Center.
We thank the anonymous referee for valuable comments and suggestions.
T.S.\ was supported by the ALMA Japan Research Grant of
NAOJ ALMA Project, NAOJ-ALMA-248.

\vspace{5mm}
\facilities{ALMA}
\software{CASA \citep{2007ASPC..376..127M}}

\bibliography{TPatm_ref}{}
\bibliographystyle{aasjournal}

\end{document}